\documentclass[11pt,a4paper]{article}
\usepackage{app/jheppub}
\usepackage{mathrsfs}
\usepackage{xcolor}
\newcommand{\sL}{\!\scriptscriptstyle L}
\newcommand{\sR}{\!\scriptscriptstyle R}

\newcommand{\PP}{\mathbb{P}}
\newcommand{\PL}{\PP_{\sL}}
\newcommand{\PR}{\PP_{\sR}}

\newcommand{\ubar}{\bar{u}}
\newcommand{\sW}{\!\scriptscriptstyle W}

\newcommand{\dbar}{\overline{d}}

\title{\boldmath 
Flavor Anomalies
Accommodated  in A Flavor Gauged Two Higgs Doublet Model}
\author{Junmou Chen, Qiaoyi Wen, Fanrong Xu, Mengchao Zhang}%\footnote{Corresponding author.}}

\affiliation%[1]
{Department of Physics and Siyuan Laboratory, Jinan University,\\
$\;$Guangzhou 510632, P.R. China}
\emailAdd{fanrongxu@jnu.edu.cn}

\abstract{
The $3.1\sigma$ $R_K$ anomaly after Moriond 2021 and $3.3 \sigma$ $\Delta a_\mu$ from Fermilab
Muon g-2 experiment implicate that the lepton flavor universality violation (LFUV) may play a role in the
exploration of  new physics. 
A Flavor Gauged Two-Higgs Doublet Model (FG2HDM) is proposed and investigated in this work.
To get rid of the 
the redundancy in Yukawa coupling of 2HDM-III, a specific U(1) flavor symmetry is introduced. 
The charge difference between two scalar doublets forbid the appearance of pseudoscalar and hence
there are only three particles (a charged and neutral heavy scalar together with a neutral gauge boson) 
adding to SM particle spectrum.
The heavy neutral scalar-mediated flavor-changing interactions occur among down-type quarks. 
With obvious difference from 2HDM-II,   the charged Higgs in FG2HDM can 
naturally explain $R_{D^{(*)}}$ anomaly.
The heavy neutral vector boson $Z'$, changing flavor for down-type quark uniquely as well, provides a solution 
to $R_{K^{(*)}}$. The  anomalous magnetic dipole moment (AMDM) of muon and electron, especially 
the new released $\Delta a_\mu$, can further discriminate $Z'$ parameter space.
}


\begin{document}
\maketitle
\flushbottom

\section{Introduction}
\label{sec:int}

So far  the Standard Model (SM) is consistent with experiments well, except some anomalies.
One type of anomalies occurs in B meson decays. In 2012 BaBar firstly measured \cite{Lees:2012xj} the ratio between
$\bar{B}\to D ^{(*)}\tau^- \bar\nu_\tau$ and   $\bar{B}\to D ^{(*)} \ell^- \bar\nu_\ell$  ($\ell = e,\mu$), 
 which exceeded SM  expectation by $2.0\sigma$ and $2.7\sigma$, respectively. This is the so-called $R_{D}$ and $R_{D^*}$ anomaly. Though there is a tension with Belle measurement in 2020 \cite{Belle:2019rba} , giving a more SM-like result, it is not the time to make a clear conclusion.
The lepton flavor universality (LFU) is expected to be satisfied in SM. In recent years, however, 
a violation of LFU (or lepton non-universality) has been unfoled  in semileptonic decay $B\to K^{(*)} \ell^+\ell^-$. In 2014, LHCb measured 
the ratio between $B^+\to K^+ \mu^+\mu^-$ and $B^+\to K^+ e^+ e^-$ and found a deviation from SM prediction 
$R_K^{\rm{SM}}=1\pm 0.01$  by $2.6 \sigma$ \cite{Bordone:2016gaq} . After the continuous updates in 2019 by LHCb \cite{Aaij:2019wad} and Belle \cite{Abdesselam:2019lab},  LHCb reported their latest result  with full Run I and Run II data during Moriond 2021\cite{Aaij:2021vac} ,
%Moriond 2021\cite{Moise:Moriond2021}
\begin{equation}
R_K=0.846^{+0.042+0.013}_{-0.039-0.012},\label{eq:RK}
\end{equation}
indicating the firm existence of $R_K$ anomaly.
Parallel to the pseudoscalar mode, this non-universality also turns up in $B\to K^*\ell^+\ell^-$ process. 
The data from LHCb in 2017 implicated a $2.2$ - $2.4\sigma$ deviation at low $q^2$ and $2.4$ - $2.5\sigma$
at central $q^2$ region  \cite{Aaij:2017vbb}, while Belle in 2019 gave a measurement more close to SM  \cite{Abdesselam:2019wac}.
Nevertheless, more precise results are anticipated in the near future with more data accumulated. 

Though involving different types of interaction, the anomalies in $b\to c \ell \nu_\ell$ and $b\to s \bar{\ell}\ell$ co-implicate that the opportunity for new physics may lie in lepton sector.  In fact,  there is a long-standing anomaly in muon anomalous magnetic dipole moment (AMDM). The SM calculation, including $\mathcal{O}(\alpha^5)$ QED and electroweak correction, NNLO hadronic vacuum polarization as well as hadronic light-by-light scattering (HLbL) contribution (see the review \cite{Aoyama:2020ynm}), differs the latest Fermilab measurement \cite{PhysRevLett.126.141801} 
 \begin{equation}
 a_\mu^{\rm{FNAL}} = (116\, 592\, 040\pm 54 )\times 10^{-11}
\end{equation}
by
%\begin{equation}
$
\Delta a_\mu = (230\pm 69)\times 10^{-11} ,
$
%\end{equation}
corresponding to a $3.3\sigma$ discrepancy. 
%Recently, the muon AMDM has been measured and its experimental value is updated 
%\begin{equation}
%a_\mu^{\rm{exp}}= \ldots
%\end{equation}
For the electron AMDM, due to an improved measurement of fine-structure constant
$\alpha$ \cite{Parker:2018vye} toward a deviation 
%\begin{equation}
$
\Delta a_e=-(8.7\pm 3.6)\times 10^{-13}
$
%\end{equation}
from theoretical prediction, corresponding to a negative $2.4\sigma$ discrepancy.
The opposite signs of AMDM for electron and muon provides an independent evidence for 
the violation of LFU. 
The latest attempt to connect B anomalies with muon AMDM
can be found in \cite{Nomura:2021oeu} after the Fermilab Muon g-2 Experiment reported their
first result.

In addition to the lepton sector, it is widely believed that 
 Physics beyond the Standard Model (BSM) is partially
encoded in the scalar sector as well. 
It is known that fermion mass, as well as the Yukawa interaction,
is co-determined by scalar VEV and Yukawa couplings. Hence a natural consequence for  extending scalar sector from its minimal model enlarges parameter space of Yukawa couplings, which can be further interpreted as one origins of LFU.
Among various multiple Higgs models, the Two-Higgs-Doublet Model (2HDM)
is one popular choice. In fact, 2HDM is contained naturally in the Minimal Supersymmetric Standard Model (MSSM), and also provides a possibility for a global U(1) symmetry leading to various axion models \cite{Kim:1986ax}. Moreover, the extra sources of CP violation in 
2HDM can also generate sufficient baryon asymmetry of the universe (BAU) which is unable 
in SM.
There are several variants of 2HDMs 
classified by the Yukawa interactions%among fermions and scalar doublets
, among which
Type I 2HDM is the simplest one as one doublet is decoupled with fermions.
In the Type II model, up-type quarks and down-type quarks couple to different doublets while charged leptons couple to the same doublet as down-type quark. 
Comparing with Type II model,
the Flipped 2HDM (or Type Y) is just to flip the doublet which the charged leptons couple to.
In another popular 2HDM, Lepton-specific 2HDM (or Type X), quarks and charged leptons are assigned to 
different scalar doublets. (For more details on 2HDMs, one can refer to the review \cite{Branco:2011iw}.) 

As pointing out in \cite{Lees:2012xj}, a tension between the 2HDM-II and  $R_{D^{(*)}}$ anomaly indicates that simple 2HDMs are challenged by current experiments. On the other hand, if  all the couplings to fermions are allowed generically, leading to Type III 2HDM \cite{Davidson:2005cw, Haber:2006ue}, the parameter space is too large to be determined. One effort is to impose the Cheng-Sher Ansatz \cite{Cheng:1987rs} to narrow the parameter space,
based on which some recent works can be found in \cite{Chen:2013qta} and works hereafter.
It is known that to open generic Yukawa coupling to all fermions bring in the dangerous flavor-changing neutral Higgs (FCNH).
In a kind of 2HDM, BGL model,  the scalar-mediated FCNC can be suppressed by small off-diagonal elements of CKM matrix under a global flavor symmetry\cite{Branco:1996bq}. Recently, a model to localize this flavor symmetry have been proposed \cite{Celis:2015ara} and  developed \cite{Ordell:2019zws,Ordell:2020yoq}, in which  a new gauge boson corresponding to the U(1) gauge group in charge of flavor symmetry and an extra scalar singlet are introduced in addition to the original 2 doublets in 2HDM. %Indeed some of the models in \cite{Celis:2015ara} can explain B anomalies.

We learnt some lessons 
from the above variant models of general 2HDM-III.
On one hand, there are too many degrees of freedom in Yukawa coupling and scalar potential. On the other hand, to restrict the 
redundant parameters, more symmetries and hence extra model dependent parameters are required. One needs to keep a balance between the "restriction" and the "freedom".
%What is the more economic way?  
In this paper, we provide a more economic solution by proposing the flavor gauged 2HDM (FG2HDM). 
The degrees of freedom in Yukawa couplings are reduced by imposing a U(1) local flavor symmetry, similar 
as the BGL models, in the price of
introducing a new neutral gauge boson with no other particles  adding to particle spectrum. 
The new gauge boson together with exotic Higgs provides a source for lepton non-universality.

This paper is organized as follows.  In Sec.\ref{sec:FG2HDM} we present the main structure
of FG2HDM.  The FG2HDM contribution to $R_K$ and $R_{K^*}$, AMDM of charged leptons
as well as $b\to c\tau\nu$ processes are calculated in Sec.\ref{sec:FA}.
%  are considered in Sec. \ref{sec:RK}. We further
%discuss FG2HDM contribution to AMDM of charged leptons and rare decay $\mu\to e\gamma$ in Sec. \ref{sec:gm2}.
%The new physics contribution to anomalies in $b\to c\tau\nu$  from charged Higgs are calculated
%in Sec. \ref{sec:RD}.
In Sec. \ref{sec:num} a combined numerical analysis is performed and solution space is given.
The conclusion and outlook are made in Sec. \ref{sec:con}. One can refer to
Appendix \ref{app:details} for more model details.

%\newpage

\section{The Flavor Gauged Two-Higgs Doublet Model}
\label{sec:FG2HDM}

The Flavor Gauged Two-Higgs Doublet Model (FG2HDM) is developed from (gauged) BGL model 
\cite{Branco:1996bq, Celis:2015ara}. Imposing the U(1) flavor symmetry on Yukawa interaction,  Yukawa couplings
have particular texture which further helps to tune the 
FCNC process mediated by neutral scalars. The U(1) charges of fermions and scalars
are assigned with different charges to satisfy anomaly-free condition. 
Especially, we do not introduce more scalar fields in addition to the 2 doublets in 2HDM. 
With specific quantum numbers of U(1), some terms in the scalar potential 
is closed comparing with the most generic one.

\subsection{Scalar sector}
\label{subsec:scalar}

The scalar potential containing two scalar doublets  in FG2HDM is of the form  
\begin{eqnarray}
V(\Phi_1,\Phi_2)&=&m_{11}^2 \Phi_1^\dagger\Phi_1 + m_{22}^2 \Phi_2^\dagger\Phi_2
+\frac{\lambda_1}{2}\left(\Phi_1^\dagger\Phi_1
\right)^2+\frac{\lambda_2}{2}\left(\Phi_2^\dagger\Phi_2
\right)^2\nonumber\\
&&+\lambda_3 \left(\Phi_1^\dagger\Phi_1\right)\left( \Phi_2^\dagger\Phi_2\right)
+\lambda_4 \left(\Phi_1^\dagger\Phi_2\right)\left(\Phi_2^\dagger\Phi_1\right),
\end{eqnarray}
where  
$SU(2)_L$ scalar doublet is notated as $\Phi_i=\left(\phi_i^+, \frac{1}{\sqrt{2}}(\rho_i+ i \eta_i +v_i)\right)^T$ and CP violating phases in VEV are not included.
%\begin{equation}
%V(\Phi_1,\Phi_2)=m_{11}^2 \Phi_1^\dagger\Phi_1 + m_{22}^2 \Phi_2^\dagger\Phi_2
%+\frac{\lambda_1}{2}\left(\Phi_1^\dagger\Phi_1
%\right)^2+\frac{\lambda_2}{2}\left(\Phi_2^\dagger\Phi_2
%\right)^2%\nonumber\\
%&&
%+\lambda_3 \left(\Phi_1^\dagger\Phi_1\right)\left( \Phi_2^\dagger\Phi_2\right)
%+\lambda_4 \left(\Phi_1^\dagger\Phi_2\right)\left(\Phi_2^\dagger\Phi_1\right),
%\end{equation}
Comparing with a more generic potential, 
the vanishment of $m_{12}$ and $\lambda_5$ terms is due to the different charges 
of new gauge group for the two doublets,\footnote{
For convenience let us adopt this scenario firstly, and
later we will show how to realize this conjecture explicitly. 
}
leading the vanishing mass for the two pseudoscalars. ( See Eq. (6) in \cite{Branco:2011iw}.)
The absence of physical pseudoscalar differs from other ordinary 2HDM in literatures. 
The mass terms for the remaining scalars are
\begin{align}
& \mathcal{L}_{\phi^\pm}=-\frac12 \lambda_4 v_1 v_2 \left(\begin{array}{cc} \phi_1^-, & \phi_2^-\end{array}\right)
\left(\begin{array}{cc}
\frac{v_2}{v_1} & -1 \\
-1& \frac{v_1}{v_2}\end{array}\right)
\left(\begin{array}{c}
\phi_1^+\\ \phi_2^+
\end{array}\right)
\nonumber\\
&\mathcal{L}_{\rho} = -\frac12\left(\begin{array}{cc} \rho_1, \rho_2 \end{array}\right)
\left(\begin{array}{cc}
\lambda_1 v_1^2 & \lambda_{34} v_1 v_2\\
 \lambda_{34} v_1 v_2 & \lambda_2 v_2^2\end{array}\right)
 \left(
 \begin{array}{c}
 \rho_1 \\ \rho_2
 \end{array}\right)
\end{align}
%%QiaoyiWen's code
%\textcolor{blue}{
%	\begin{equation}
%	\mathcal{L}_{\phi^\pm}=-\frac{1}{2}\lambda_4 v_1 v_2 \left(\begin{array}{cc} \phi_1^-, & \phi_2^-\end{array}\right)
%	\left(\begin{array}{cc}
%		\frac{v_2}{v_1} & -1 \\
%		-1& \frac{v_1}{v_2}\end{array}\right)
%	\left(\begin{array}{c}
%		\phi_1^+\\ \phi_2^+
%	\end{array}\right)\nonumber
%\end{equation}
%\begin{equation}
%	\mathcal{L}_{\rho} = -\frac{1}{2}\left(\begin{array}{cc} \rho_1, \rho_2 \end{array}\right)
%	\left(\begin{array}{cc}
%		\lambda_1 v_1^2 & \lambda_{34} v_1 v_2\\
%		\lambda_{34} v_1 v_2 & \lambda_2 v_2^2\end{array}\right)
%	\left(
%	\begin{array}{c}
%		\rho_1 \\ \rho_2
%	\end{array}\right)\nonumber
%\end{equation}}
%\textcolor{blue}{
%$$\tan\alpha=\frac{\lambda_{2}\sin^2\beta-\lambda_{1}\cos^2\beta-\sqrt{(\lambda_{1}\cos^2\beta-%\lambda_{2}\sin^2\beta)^2+4\lambda^2_{34}\sin^2\beta\cos^2\beta}}{2\lambda_{34}\sin\beta\cos\beta}$$
%	$$\rho_{\rm eigen value }=\frac{1}{2}v^2\Big[\lambda_{1}\cos^2\beta+\lambda_{2}\sin^2\beta\pm\sqrt{(\lambda_1\cos^2\beta-\lambda_2\sin^2\beta)^2+4\lambda_{34}^2\sin^2\beta\cos^2\beta}\Big]$$
	%}
%%
%
where $\lambda_{34}=\lambda_3+\lambda_4$.
%We may find that the mass matrix for $\eta_{1,2}$ automatically vanishes for 
%the forbidding of $m_{12}$ and $\lambda_5$.
After diagnolization, another massless charged scalar, together with the two
massless neutral pseudoscalars,  plays the role of Goldstone bosons
which give masses to massive gauge bosons, $W^+, Z$ and $Z'$. 
The rotational matrices, transforming 
scalars from gauge eigenstates to mass eigenstates, are in the convention of 
\begin{align}
&\left(\begin{array}{c} h \\ H^0\end{array}\right)=
\left(\begin{array}{cc}
\cos\alpha & \sin\alpha \\ -\sin\alpha & \cos\alpha\end{array}\right)
\left(\begin{array}{c} \rho_1 \\ \rho_2\end{array}\right),
\nonumber\\
&\left(\begin{array}{c} G^\pm \\ H^\pm\end{array}\right)=
\left(\begin{array}{cc}
\cos\beta & \sin\beta \\ -\sin\beta & \cos\beta\end{array}\right)
\left(\begin{array}{c} \phi^\pm_1 \\ \phi^\pm_2\end{array}\right).
\end{align}
with  $\tan\beta=\frac{v_2}{v_1}$, vacuum expected value $v=\sqrt{v_1^2+v_2^2}=(\sqrt{2}G_F)^{-\frac12}\approx 246\,\textrm{GeV}$
and 
\begin{equation*}
\tan\alpha=\frac{1}{2\lambda_{34}\sin\beta\cos\beta}\left[
\lambda_{2}\sin^2\beta-\lambda_{1}\cos^2\beta-\sqrt{(\lambda_{1}\cos^2\beta-\lambda_{2}\sin^2\beta)^2+4\lambda^2_{34}\sin^2\beta\cos^2\beta}
\right],
\end{equation*}
which is governed by Higgs coupling $\lambda_{1,2,3,4}$ and $\beta$.
Note the angles $\alpha$ and $\beta$ are defined in the rotation of neutral scalar and charged scalar, respectively.
In the limit of $\cos(\beta-\alpha)\to 1$ (or equivalently $\sin(\beta-\alpha)\to 0$), the mass basis (the basis we adopt here) is identical to Higgs basis.
For the charged scalar $G^{\pm}$ absorbed by $W^{\pm}$, there remain three physical scalars:
the discovered  $125\,\textrm{GeV}$ neutral scalar $h$, 
the undiscovered exotic heavy neutral scalar  $H^0$ and heavy charged scalar $H^+$.
%\subsubsection*{Scalar interaction}
%\textcolor{blue}{
%The interaction Lagrangian for scalars after electroweak symmetry breaking is:
%\begin{equation}
%	\begin{aligned}
%	-\mathcal{L}_\text{scalar}&=\frac{\lambda_{1}}{2}\Big(\phi_1^+\phi_1^-\phi_1^+\phi_1^-+\frac14\rho_1^4+2v_1\rho_1\phi_1^+\phi_1^-+\rho_1^2\phi_1^+\phi_1^-+v_1\rho_1^3\Big)\\
%	&+\frac{\lambda_{2}}{2}\Big(\phi_2^+\phi_2^-\phi_2^+\phi_2^-+\frac14\rho_2^4+2v_2\rho_2\phi_2^+\phi_2^-+\rho_2^2\phi_2^+\phi_2^-+v_2\rho_2^3\Big)\\
%	&+\lambda_3\Big(v_2\rho_2\phi_1^+\phi_1^-+v_1\rho_1\phi_2^+\phi_2^-+\frac12 v_2\rho_2\rho_1^2+\frac12 v_1\rho_1\rho_2^2\Big)\\
%	&+\lambda_3\Big(\phi_1^+\phi_1^-\phi_2^+\phi_2^-+\frac12\rho_2^2\phi_1^+\phi_1^-+\frac12\rho_1^2\phi_2^+\phi_2^-+\frac14\rho_1^2\rho_2^2\Big)\\
%	&+\lambda_4\Big(\phi_1^+\phi_1^-\phi_2^+\phi_2^-+\frac12v_1\rho_1\rho_2^2+\frac12v_2\rho_2\rho_1^2+\frac14\rho_1^2\rho_2^2\Big)\\
%	&+\lambda_4\Big(\frac12\rho_1\rho_2\phi_2^+\phi_1^-+\frac12v_1\rho_2\phi_2^+\phi_1^-+\frac12v_2\rho_1\phi_2^+\phi_1^-+h.c\Big)
%	\end{aligned}
%\end{equation}
%}
%\textcolor{blue}{
Without loss of generality, the interactions among scalar eigenstates are given as
%Rotate the interacting eigenstates to mass eigenstates, we obtain:
\begin{equation}
	\begin{aligned}
	-\mathcal{L}_\text{scalar}&=\frac{\lambda_{h^3}}{3!}h^3+\frac{\lambda_{h^2H^{0}}}{2}h^2H^{0}+\frac{\lambda_{hH^{0^2}}}{2}hH^{0^2 }+\frac{\lambda_{H^{0^3}}}{3!}H^{0^3}\\
	&+\lambda_{hH^+H^-}hH^+H^-+\lambda_{HH^+H^-}HH^+H^-+\frac{\lambda_{H^+H^-H^+H^-}}{4}H^+H^-H^+H^-\\
	&+\frac{\lambda_{h^4}}{4!}h^4+\frac{\lambda_{h^3H^0}}{3!}h^3H^0+\frac{\lambda_{h^2H^{0^2}}}{4}h^2H^{0^2}+\frac{\lambda_{hH^{0^3}}}{3!}hH^{0^3}+\frac{\lambda_{H^{0^4}}}{4!}H^{0^4}\\
	&+\frac{\lambda_{h^2H^+H^-}}{2}h^2H^+H^-+\lambda_{hH^0H^+H^-}hH^0H^+H^-+\frac{\lambda_{H^{0^2}H^+H^-}}{2}H^{0^2}H^+H^-
	\end{aligned}
\end{equation}
with $h^{[n]}H^{0^{[m]}}\equiv \overbrace{h\cdots h}^{n} \underbrace{H^0\cdots H^0}_{m}$ and relavent 
couplings can be found in Appendix \ref{app:scalar}.
%}

\subsection{Yukawa interaction}

The Yukawa interaction, including both quark and lepton sectors, are generally  
in the form of
\begin{eqnarray}
-\mathcal{L}_Y&=& \overline{Q^0_L}(Y_1^d\Phi_1+Y_2^d\Phi_2)d_R^0
+ \overline{Q^0_L}(Y_1^u\tilde{\Phi}_1+Y_2^u\tilde{\Phi}_2)u_R^0\nonumber\\
&&+ \overline{L^0_L}(Y_1^\ell{\Phi_1}+Y_2^\ell{\Phi_2})e_R^0
+ \overline{L^0_L}(Y_1^\nu\tilde{\Phi}_1+Y_2^\nu\tilde{\Phi}_2)\nu_R^0
+h.c.
\end{eqnarray}
in which the fermion fields with superscirpt $0$ denotes the fields in gauge eigenstate.
In total there are eight $3\times 3$ Yukawa matrices. In current work, we assume neutrino mass is Dirac type generated by the corresponding Yukawa matrices $Y_i^\nu$.

\subsubsection*{Fermion mass}

After spontaneous symmetry breaking, fermion mass terms can be written as
\begin{equation}
-\mathcal{L}_m=\bar{u}_L M_u u_R+ \bar{d}_L M_d d_R
+\bar{\ell}_L M_\ell \ell_R
+\bar{\nu}_L M_\nu \nu_R+h.c.~,
\end{equation}
where $M_f$ ($f=u,d,\ell,\nu$) is diagonal mass matrix of fermions, rotated from $\tilde{M}_f$,
\begin{equation}
 M_f= U_{fL}^\dagger \tilde{M}_f U_{fR}, \qquad
 \tilde{M}_f = \frac{1}{\sqrt{2}}(v_1 Y_1^f + v_2 Y_2^f),
\end{equation}
and $U_{fL(R)}$ are the rotation matrices connected the fermions in mass eigenstate and weak 
eigenstate via
\begin{equation}
f^0_{L(R)}=U_{fL(R)} f_{L(R)}.
\end{equation}
The CKM and PMNS matrices then can be defined as
\begin{equation}
V\equiv V_{\rm CKM} = U_{uL}^\dagger U_{dL},\quad
U\equiv V_{\rm PMNS} = U_{\nu L}^\dagger U_{\ell L},
\end{equation}
which will play a role in gauge interactions and fermion scalar interactions.

\subsubsection*{The interactions among fermions and scalars}

From the analysis in scalar potential sector, there are three physical scalar particles ($h, H^0, H^+$) in the FG2HDM as the others are eaten by gauge bosons. 
After rotation and redefinition, one may write down
the interaction among fermions and physical scalars in mass eigenstate as follows,
%\begin{eqnarray}
%-\mathcal{L} &=& \frac{\sqrt{2}}{v} H^+  \left[ \bar{u}\left(V_{\rm CKM} N_d \PR - N_u^\dagger V_{\rm CKM} \PL\right)d 
%+ \bar{\nu}\left(V_{\rm PMNS} N_\ell \PR - N_\nu^\dagger V_{\rm PMNS} \PL\right)\ell
%\right]
% \nonumber\\
%&&+\frac{1}{v}  H^{\rm{SM}} [\ubar M_u u + \dbar M_d d +\bar{\ell} M_\ell \ell  +\bar{\nu} M_\nu \nu ] \nonumber\\
%&& +\frac{1}{v}  H^{0} [\ubar N_u u + \dbar N_d d +\bar{\ell} N_\ell \ell  +\bar{\nu} N_\nu \nu ]+h.c.
%\end{eqnarray}

%%QiaoyiWen's code
%\textcolor{orange}{
%\begin{eqnarray}
%	-\mathcal{L} &=& \frac{\sqrt{2}}{v} H^+  \left[ \bar{u}\left(V_{\rm CKM} N_d \PR - N_u^\dagger V_{\rm CKM} %\PL\right)d 
%	+ \bar{\nu}\left(V_{\rm PMNS} N_\ell \PR - N_\nu^\dagger V_{\rm PMNS} \PL\right)\ell
%	\right]
%	\nonumber\\
%	&& +\frac{1}{v}  H [\ubar N_u u + \dbar N_d d +\bar{\ell} N_\ell \ell  +\bar{\nu} N_\nu \nu ]+h.c.\nonumber\\
%	&&+\frac{1}{v}  H^{\rm{SM}} [\ubar M_u u + \dbar M_d d +\bar{\ell} M_\ell \ell  +\bar{\nu} M_\nu \nu ] \nonumber
%\end{eqnarray}
%}
%\textcolor{blue}{
	\begin{eqnarray}
		-\mathcal{L} &=& \frac{\sqrt{2}}{v} H^+  \left[ \bar{u}\left(V_{\rm CKM} N_d \PR - N_u^\dagger V_{\rm CKM} \PL\right)d 
		+ \bar{\nu}\left(V_{\rm PMNS} N_\ell \PR - N_\nu^\dagger V_{\rm PMNS} \PL\right)\ell
		\right]+h.c.
		\nonumber\\
		&& +\frac{1}{v}  \left[\cos(\beta-\alpha)H^0-\sin(\beta-\alpha)h\right] [\ubar N_u u + \dbar N_d d +\bar{\ell} N_\ell \ell  +\bar{\nu} N_\nu \nu ]\nonumber\\
		&&+\frac{1}{v}  \left[\sin(\beta-\alpha)H^0+\cos(\beta-\alpha)h\right] [\ubar M_u u + \dbar M_d d +\bar{\ell} M_\ell \ell  +\bar{\nu} M_\nu \nu ] \label{eq:Yukawa}
	\end{eqnarray}
%}
%%
where the diagonal mass matrices are
 of the forms $M_u={\textrm{diag}}(m_u, m_c, m_t), M_d={\textrm{diag}}(m_d, m_s, m_b),$ 
 $M_\ell={\textrm{diag}}(m_e, m_\mu, m_\tau), M_\nu={\textrm{diag}}(m_{\nu_1}, m_{\nu_2}, m_{\nu_3})$, explicitly. 
 Note in Eq. (\ref{eq:Yukawa}) all the fermion are in mass eigenstate, including neutrinos.
Especially, according to the field in original Lagrangian, the tree level flavor-changing current induced by scalar
and controlled by $N_f$, given,
\begin{equation}
N_f=\frac{1}{\sqrt{2}}U_{fL}^\dagger(v_1 Y_2^f - v_2 Y_1^f) U_{fR}.
\end{equation}
Apparently,  the form of $N_f$ is determined by the choice of $Y_i^f$, which further can be regarded as
a result of new symmetry. Without loss of generality, $N_f$
 can also be simplified to %either
 \begin{equation}
 N_f= -\frac{v_2}{v_1} M_f + \frac{v_2}{\sqrt{2}} \left(\frac{v_2}{v_1} + \frac{v_1}{v_2} \right)
 U_{fL}^\dagger
 Y_2^f U_{fR}.
 \end{equation}
The Yukawa coupling matrices, so far, have not received any restrictions and hence have the most general structure.
We will show in below, under some particular symmetries, the special structure of Yukawa will bring in the good features:
the controlled FCNC and the connection to CKM matrix.

% or 
 %\begin{equation}
% N_f= \frac{v_1}{v_2} M_f - \frac{v_1}{\sqrt{2}} \left(\frac{v_2}{v_1} + \frac{v_1}{v_2} \right)
 %U_{fL}^\dagger
% Y_1^f U_{fR},
% \end{equation}
%in different situations.

\subsubsection*{A specific texture for Yukawa coupling matrices}
Under the U(1) symmetry assigned with the particular 
quantum number shown in Appendix \ref{app:symmetry}, 
the Yukawa matrices are of special forms and 
the tree-level FCNC can be tuned by CKM matrix. 
Suppose the forms of quark Yukawa textures are
\begin{equation}
Y_1^u= \left(\begin{array}{ccc}
* & * & 0 \\
* & * & 0 \\
0 & 0 & 0 
\end{array}\right),\quad
Y_2^u= \left(\begin{array}{ccc}
0 & 0 & 0 \\
0 & 0 & 0 \\
0 & 0 & * 
\end{array}\right),\quad
Y_1^d= \left(\begin{array}{ccc}
* & * & * \\
* & * & * \\
0 & 0 & 0 
\end{array}\right),\quad
Y_2^d= \left(\begin{array}{ccc}
0 & 0 & 0 \\
0 & 0 & 0 \\
* & * & * 
\end{array}\right),\label{eq:quarkYukawa}
\end{equation}
in which `$*$' denotes a non-zero arbitrary number in corresponding entry. 
Combing the definitions of $\tilde{M}_f$, $N_f$, one  easily obtains the coupling matrix for quarks and scalar
\begin{align}
& N_u = -\frac{v_2}{v_1} \,{\rm diag}(m_u, m_c, 0) + \frac{v_1}{v_2} \,  {\rm diag}(0, 0, m_t), \nonumber\\
& (N_d)_{ij} =  -\frac{v_2}{v_1} (M_d)_{ij} +\left(\frac{v_2}{v_1}+\frac{v_1}{v_2}\right) V_{i3}^\dagger V_{3j} (M_d)_{jj},  
 \end{align}
 for the quark sector.
 The Yukawa texture for leptons are of the form
\begin{equation}
Y_1^\ell= \left(\begin{array}{ccc}
0 & 0 & 0 \\
0 & * & 0 \\
0 & 0 & * 
\end{array}\right),\quad
Y_2^\ell= \left(\begin{array}{ccc}
* & 0 & 0 \\
0 & 0 & 0 \\
0 & 0 & 0 
\end{array}\right),\quad
Y_1^\nu= \left(\begin{array}{ccc}
0 & 0 & 0\\
0 & 0 & 0 \\
* & * & * 
\end{array}\right),\quad
Y_2^\nu= 0, \label{eq:lepYukawa}
\end{equation}
 thus the  couplings among leptons and scalar are
 \begin{align}
& N_\nu = -\frac{v_2}{v_1} M_\nu,\nonumber \\
%& N_\ell = -\frac{v_2}{v_1}{\textrm{diag}}(m_e,m_\mu,0)+\frac{v_1}{v_2}{\textrm{diag}}(0,0,m_\tau)    
& N_\ell = -\frac{v_2}{v_1}{\textrm{diag}}(0,m_\mu,m_\tau)+\frac{v_1}{v_2}{\textrm{diag}}(m_e,0,0)    
\end{align}
for leptons.
 
Among all the fermions, only down-type quark receives FCNC mediated by neutral Higgs.

 %are defined as diagonalized  mass matrices.

% \vspace{3cm}
 
 %\newpage

%\newpage

\subsection{Gauge interaction}

The kinematic terms for scalar fields in Lagrangian is
\begin{equation}
\mathcal{L}_G = (D_\mu \Phi_1)^\dagger (D^\mu \Phi_1)+ (D_\mu \Phi_2)^\dagger (D^\mu \Phi_2),
\label{eq:LG}
\end{equation}
in which the gauge derivative for scalar field is defined as
\begin{equation}
D_\mu \Phi_j =\left(\partial_\mu - i g_1 Y_j B_\mu - i g' Q_j \hat{Z}'_\mu -ig_2 \frac{\vec{\tau}}{2} \cdot \vec{W}_\mu \right)
\Phi_j,
\end{equation}
and hypercharge under $U(1)_L$ is known as $Y_1=Y_2=\frac12$, $Q_j$ is quantum number of $\Phi_j$ under
$U(1)'$, given in Eq. (\ref{eq:charge}).

\subsubsection*{Gauge boson mass}
After spontaneous symmetry breaking, the mass terms for gauge bosons are
\begin{equation}
\mathcal{L}_m^G= \left(\begin{array} {ccc}
B& W^3 & \hat{Z}' \end{array}\right) \tilde{M} \left(\begin{array}{c}
B \\ W^3 \\ \hat{Z}' \end{array}\right) =\left(\begin{array} {ccc}
A& Z & Z' \end{array}\right) {M}_{d} \left(\begin{array}{c}
A \\ Z \\ Z' \end{array}\right).
\end{equation}
%in which
%{\scriptsize{
%\begin{eqnarray}
%\tilde{M}&=&\left(\begin{array}{ccc}
%\frac18 g_1^2 v^2 & -\frac18 g_1g_2 v^2  & \frac14 g_1 g' v^2 (Q_1 \cos^2\beta + Q_2 \sin^2\beta) \\
%-\frac18 g_1 g_2 v^2 & \frac18 g_2^2 v^2 & -\frac14 g_2 g' v^2 (Q_1 \cos^2\beta + Q_2 \sin^2\beta)\\
%\frac14 g_1 g' v^2 (Q_1 \cos^2\beta + Q_2 \sin^2\beta) & -\frac14 g_2 g' v^2 (Q_1 \cos^2\beta + Q_2 \sin^2\beta) &
%\frac12 g^{'2}v^2 (Q_1^2 \cos^2\beta + Q_2^2 \sin^2\beta)
%\end{array}
%\right)\\
%&=& \frac18 (g_1^2 + g_2^2) v^2 
%\left(\begin{array}{ccc}
%\sin^2\theta_W & -\sin\theta_W\cos\theta_W  & 2\sin\theta_W\sin\xi (Q_1 \cos^2\beta + Q_2 \sin^2\beta) \\
 %-\sin\theta_W\cos\theta_W& \cos^2\theta_W & -2\cos\theta_W \sin\xi (Q_1 \cos^2\beta + Q_2 \sin^2\beta)\\
% 2\sin\theta_W\sin\xi (Q_1 \cos^2\beta + Q_2 \sin^2\beta)  &-2\cos\theta_W \sin\xi (Q_1 \cos^2\beta + Q_2 \sin^2\beta)&
%4\sin^2\xi (Q_1^2 \cos^2\beta + Q_2^2 \sin^2\beta)
%\end{array}
%\right)\nonumber
%\end{eqnarray}
%}}
%and $\sin\theta_W=\frac{g_1}{\sqrt{g_1^2+g_2^2}}$, $\sin\xi= \frac{g'}{\sqrt{g_1^2+g_2^2}}$.
Impose the condition between the charges of two scalar doublet
\begin{equation}
Q_1=-Q_2 \tan^2\beta,
\label{eq:rela-phi}
\end{equation}
the mixing of $Z'$ and other two gauge bosons is decoupled, and the mass matrix can be simplified as
\begin{equation}
\tilde{M} = 
\frac18 (g_1^2 + g_2^2) v^2 
\left(\begin{array}{ccc}
\sin^2\theta_W & -\sin\theta_W\cos\theta_W  & 0 \\
 -\sin\theta_W\cos\theta_W& \cos^2\theta_W &0\\
 0 &0 &
4\sin^2\xi (Q_1^2 \cos^2\beta + Q_2^2 \sin^2\beta)
\end{array}
\right)
\end{equation}
with $\sin\theta_W=\frac{g_1}{\sqrt{g_1^2+g_2^2}}$, $\sin\xi= \frac{g'}{\sqrt{g_1^2+g_2^2}}$.
Then the diagonalized mass matrix is
\begin{equation}
M_d=\frac18 (g_1^2 + g_2^2) v^2 
\left(\begin{array}{ccc}
0 &  & \\
 & 1 & \\
  & &
4 Q_2^2\sin^2\xi  \tan^2\beta
\end{array}
\right).
\end{equation}
For more explicit, we obtain the relation of masses between $Z'$ and $Z$,
\begin{equation}
\frac{m_{Z'}}{m_Z}= 2Q_2 \sin\xi \tan\beta,
\end{equation}
which is determined by $Q_2$, $\xi$ and $\beta$.
And the rotation matrix connected mass eigenstate and gauge eigenstate is
\begin{equation}
\left(\begin{array}{c}
A \\ Z \\ {Z}' \end{array}\right)
= U  \left(\begin{array} {c}
B\\ W^3 \\ \hat{Z}' \end{array}\right),\qquad
U=\left(\begin{array}{ccc}
\cos\theta_W & \sin\theta_W & 0 \\
-\sin\theta_W & \cos\theta_W & 0 \\
0& 0 & 1
\end{array}\right).
\end{equation}

Note the kinetic mixing between two gauge bosons of SM $U(1)$ and $U(1)'$ is not forbidden in principle.
For the convenience we choose to close the kinetic mixing in current work.  

\subsubsection*{The interaction among gauge bosons and scalars}

The three-point interaction among gauge boson and scalars can be extracted from
full expansion of Eq. (\ref{eq:LG}), giving
\begin{equation}
	\begin{aligned}
\mathcal{L} =&  -i\frac{e}{2 s_{\sW} c_{\sW}}( s_{2\sW} A_\mu + c_{2\sW}  Z_\mu) (\partial^\mu H^- H^+ -\partial^\mu H^+H^-) \\
&-i g'(Q_1 \sin^2\beta +Q_2 \cos^2 \beta) Z'_\mu  (\partial^\mu H^- H^+ -\partial^\mu H^+H^-)\nonumber\\
&-i\frac12 g_2\sin(\alpha-\beta)[\partial^\mu h(W_\mu^-H^+ - W_\mu^+H^-)+h(\partial^\mu H^-W_\mu^+-\partial^\mu H^+W_\mu^-)]\\
&-i\frac12 g_2\cos(\alpha-\beta)[\partial^\mu H^0(W_\mu^-H^+ - W_\mu^+H^-)+H^0(\partial^\mu H^-W_\mu^+-\partial^\mu H^+W_\mu^-)]\\
&+\frac12 g_2g' v\sin2\beta(Q_2-Q_1)Z'^\mu(W_\mu^+H^-+W_\mu^-H^+)+\frac12 g_2^2 v W_\mu^+W^{-\mu}[\cos(\beta-\alpha)h+\sin(\beta-\alpha)H^0]\\
&+g'^{2} v (Q_1^2\cos\beta\cos\alpha+Q_2^2\sin\beta\sin\alpha) Z^{'}_\mu Z^{'\mu} h+\frac{g'e v}{s_{\sW} c_{\sW}} (Q_1\cos\beta\cos\alpha+Q_2\sin\beta\sin\alpha) Z'_\mu Z^\mu h\\
&+g'^{2} v   (Q_2^2\sin\beta\cos\alpha-Q_1^2\cos\beta\sin\alpha) Z^{'}_\mu Z^{'\mu} H^0+\frac{g'e v}{s_{\sW} c_{\sW}}(Q_2\sin\beta\cos\alpha-Q_1\cos\beta\sin\alpha) Z'_\mu Z^\mu  H^0\\
&+\frac{e^2 v}{4 s_{\sW}^2 c_{\sW}^2}[\cos(\beta-\alpha)h+\sin(\beta-\alpha) H^0]Z_\mu Z^\mu
	\end{aligned}
\end{equation}
%}

%\vspace{3cm}
%\newpage

\subsection{ Fermion current under new gauge symmetry }
By imposing the relation between two charges of scalar doublet Eq. (\ref{eq:rela-phi}), 
the mixing of $Z'$ to other neutral gauge bosons is decoupled. 
Hence all the currents among fermions and gauge bosons in SM keep unchanged, while
the new current brought by $Z'$ is given as
\begin{equation}
 \mathcal{L}_{Z'}=g'   \bar{f} (Q_{fL} \gamma^\mu \PL + Q_{fR} \gamma^\mu \PR) f Z'_\mu
\label{eq:fZprime}
\end{equation}
with $f=u, d, \nu, \ell$. The coupling to fermions with different chirality is governed by different $U(1)'$ charges, which in
general are given as
\begin{equation}
Q_{fL}=U_{fL}^\dagger X_{f_L} U_{fL},\qquad
Q_{fR}=U_{fR}^\dagger X_{f_R} U_{fR}.
\end{equation}
We can easily  have the explicit expressions for different types of fermions,
\begin{align}
& Q_{uL}=\frac12\left(\begin{array}{ ccc}
Q_{u_R}+Q_{d_R} & &\\
& Q_{u_R}+Q_{d_R} & \\
 & & Q_{t_R}+Q_{d_R} \end{array}\right), \label{eq:Zcharge} \\
&  Q_{dL}=
\frac12 (Q_{u_R}+Q_{d_R})\left(\begin{array}{ ccc}
1 & &\\
& 1 & \\
 & & 1 \end{array}\right)+
\frac12 (Q_{t_R}-Q_{u_R})\left(\begin{array}{ ccc}
|c_1|^2 &c_1^* c_2 & c_1^* c_3\\
c_2^* c_1& |c_2|^2 & c_2^* c_3 \\
 c_3^* c_1&c_3^* c_2 & |c_3|^2  \end{array}\right), \nonumber\\
& Q_{uR}=
%Q_{u_R}\left(\begin{array}{ ccc}
%1 & &\\
%& 1 & \\
% & & 1 \end{array}\right) +
%(Q_{t_R}-Q_{u_R})\left(\begin{array}{ ccc}
%0 &  &\\
%& 0 & \\
% & & 1  \end{array}\right)
% =
\left(\begin{array}{ ccc}
Q_{u_R} & &\\
& Q_{u_R} & \\
 & & Q_{t_R} \end{array}\right),\quad 
 %\nonumber \\
 Q_{dR}=
Q_{d_R}\left(\begin{array}{ ccc}
1 & &\\
& 1 & \\
 & & 1 \end{array}\right), \nonumber\\
&  Q_{\ell L}=\left(
\begin{array}{ccc}
Q_{e_L} & & \\
 & Q_{\mu_L} &  \\
 & & Q_{\tau_L}\end{array}
\right),%\nonumber\\
\quad
Q_{\ell R}=\left(
\begin{array}{ccc}
Q_{e_R} & & \\
 & Q_{\mu_R} &  \\
 & & Q_{\tau_R}\end{array}
\right),\nonumber\\
& \big(Q_{\nu L}\big)_{ij} = Q_{e_L} V^*_{ei} V_{ej} + Q_{\mu_L} V^*_{\mu i} V_{\mu j} 
+ Q_{\tau_L} V^*_{\tau i} V_{\tau j}, \quad i,j =1,2, 3, \nonumber\\
&Q_{\nu R}=0\nonumber
\end{align}
in which we have defined  $c_i\equiv V_{ti}, i=d,s,b$.
%and $d\equiv U_{u_R}|_{33}$. 
To obtain the above charges coupling to
$Z'$, we have made use of the structure of $U_{u_L}$ and $U_{u_R}$
\begin{equation}
U_{u_L}=\left(\begin{array}{ ccc}
* & * & 0\\
* & * & 0 \\
 0&0 & 1 \end{array}\right),\qquad
 U_{u_R}=\left(\begin{array}{ ccc}
* & * & 0\\
* & * & 0 \\
 0&0 & 1 \end{array}\right),
\end{equation}
%\begin{equation}
%U_{e_L}=\left(\begin{array}{ ccc}
%* & 0& 0\\
%0 & * & 0 \\
 %0&0 & * \end{array}\right),\qquad
% U_{e_R}=\left(\begin{array}{ ccc}
%* & 0 & 0\\
%0 & * & 0 \\
% 0&0 & * \end{array}\right),
%\end{equation}
and
\begin{equation}
U_{\ell_L}=U_{\ell_R}=1
\end{equation}
which are required by the form of mass matrices. 
  Apparently, FCNC occurs only in down-type quark sector
with left handed chirality. 

\section{Flavor Anomalies in FG2HDM}
\label{sec:FA}
As illustrated in above section, there are only three additional particles besides SM particles.
The FG2HDM provides a economic solution to the anomalies in $B\to D^{(*)} \tau \nu$, 
$B\to K^{(*)} \ell\bar{\ell}$ and anomalous magnetic dipole moments of muon and electron.
In this section, we explicitly calculate the characterized quantities of $R_{D^{(*)}}$ contributed from the charged Higgs, together with $R_{K^{(*)}}$ 
and $\Delta a_{\ell}$ originated from the exotic neutral gauge boson.

\subsection{ $R_{K^{(*)}}$ and $Z'$}
\label{subsec:RK}

At tree-level, the FCNC process in FG2HDM 
can be mediated by both exotic neutral scalar $H^0$ and $Z'$  in down-type quark decays. 
It is known that scalar operators do not contribute to $B\to K \ell\bar{\ell}$ process \cite{Bobeth:2007dw}, 
hence we only consider
the NP effect from $Z'$.

%\subsubsection{The new contribution in FG2HDM from $Z'$}
We adopt the following convention to describe $b\to s$ transition, in which the effective Hamiltonian  is in the form of
\begin{equation}
\mathscr{H}=-\frac{4G_F}{\sqrt{2}}V_{tb}V^*_{ts}\frac{e^2}{16\pi^2}\sum_i\left(C_i \mathcal{O}_i
+C'_i\mathcal{O}'_i\right)+h.c.
\end{equation}
where the effective operators are defined as
\begin{align}
& \mathcal{O}_7=\frac{m_b}{e}(\bar{s}\sigma_{\mu\nu}\PR b)F^{\mu\nu},\quad\;
\mathcal{O}'_7=\frac{m_b}{e}(\bar{s}\sigma_{\mu\nu}\PL b)F^{\mu\nu},\\
& \mathcal{O}_9=(\bar{s}\gamma_\mu\PL b)(\bar{\ell}\gamma^\mu \ell),\quad\quad\;\;
 \mathcal{O}'_9=(\bar{s}\gamma_\mu\PR b)(\bar{\ell}\gamma^\mu \ell),\nonumber\\
& \mathcal{O}_{10}=(\bar{s}\gamma_\mu\PL b)(\bar{\ell}\gamma^\mu\gamma_5 \ell),\quad\;
 \mathcal{O}'_{10}=(\bar{s}\gamma_\mu\PR b)(\bar{\ell}\gamma^\mu \gamma_5\ell),\nonumber\\
& \mathcal{O}_{S}=m_b(\bar{s}\PR b)(\bar{\ell}  \ell),\qquad\quad\;
\mathcal{O}'_{S}=m_b(\bar{s}\PL b)(\bar{\ell}  \ell),\;\nonumber\\
& \mathcal{O}_{P}=m_b(\bar{s}\PR b)(\bar{\ell} \gamma_5 \ell),\qquad\;\;
\mathcal{O}'_{P}=m_b(\bar{s}\PL b)(\bar{\ell} \gamma_5 \ell).\nonumber
\end{align}
The scattering amplitude for $b\to s\ell\bar{\ell}$ from $Z'$-induced FCNC in FG2HDM, which 
occurs uniquely in down-type quark sector, is
\begin{equation}
\mathcal{M} =- \frac{g^{'2}}{m_{Z'}^2} \frac14 V_{tb}V^*_{ts}\ (Q_{t_R}-Q_{u_R}) %c_2^* c_3 %Q_{\ell_L}
\left[ (Q_{\ell_R}+Q_{\ell_L})
(\bar{s}\gamma^\mu \PL b)(\bar{\ell}\gamma_\mu  \ell)+
 (Q_{\ell_R}-Q_{\ell_L})
(\bar{s}\gamma^\mu \PL b)(\bar{\ell}\gamma_\mu \gamma_5 \ell)
\right].
\end{equation}
Only the coefficients of $\mathcal{O}_{9,10}$ are corrected, hence we  extract the
modification to Wilson coefficients, giving
\begin{align}
&\Delta C_9^\ell =\frac{1}{N}\frac{g^{'2}}{m_{Z'}^2} \frac14 V^*_{ts}V_{tb}(Q_{t_R}-Q_{u_R})  (Q_{\ell_R}+Q_{\ell_L})
%c_2^* c_3,
,\\
& \Delta C_{10}^\ell = \frac{1}{N}\frac{g^{'2}}{m_{Z'}^2} \frac14 V^*_{ts}V_{tb} (Q_{t_R}-Q_{u_R})  (Q_{\ell_R}-Q_{\ell_L}),
%c_2^* c_3,
%\qquad (\textrm{sign\; issue})
\nonumber
\end{align}
with $N=\frac{4G_F}{\sqrt{2}}V_{tb}V^*_{ts}\frac{e^2}{16\pi^2}$. 
We can see $\Delta C_{9,10}^\ell$ is indeed flavor dependent and corresponding factors
can be found in Appendix \ref{app:symmetry}. It is worthy pointing out that the two degrees of freedom 
for U(1) charges in FG2HDM, formally giving
($\Delta C_9^\mu \sim \frac{11}{6} Q_{\mu_R}-Q_{d_R},
\Delta C_{10}^\mu \sim \frac16 Q_{\mu_R} + Q_{d_R},
\Delta C_9^e \sim \frac{1}{2} Q_{\mu_R}+3Q_{d_R},
\Delta C_{10}^e \sim \frac16 Q_{\mu_R} + Q_{d_R}
$), provide a chance to explain lepton flavor dependent $R_{K^{(*)}}$ anomaly.

%\section{Anomalous Magnetic Dipole Moment of Charged Lepton}
\subsection{ $(g-2)_\ell$ and $Z'$ }
\label{subsec:gm2}

%\subsubsection{FG2HDM contribution }
The anomalous
magnetic dipole moment (AMDM) of charged leptons, especially for muon and electron, are
generally taken as a platform for checking new physics associated with lepton sector. 
The flavor-conserving interaction among leptons and exotic neutral gauge boson $Z'$ in FG2HDM, according to Eq. (\ref{eq:fZprime}) and (\ref{eq:Zcharge}), indicates that 
AMDM of charged lepton can be generated via one-loop correction since $Z'$ decouples 
with photon and $Z$ in current scenario of FG2HDM.

The calculation for  a general $Z'$ contribution to AMDM at one-loop level in Feynman gauge can be found in \cite{Lynch:2001zr}. Ignoring the unphysical scalar contribution safely in the heavy mass limit of vector boson, we have $Z'$ contribution to charged lepton AMDM, denoted as
\begin{equation}
\Delta a_{\ell}=-\frac{m_{\ell}^2}{8 \pi^{2}} \frac{g'^{2}}{m_{Z'}^{2}} \frac{2}{3}\left[\left(Q_{\ell_L}^{2}+Q_{\ell_R}^{2}\right)-3 Q_{\ell_L} Q_{\ell_R}\right],
\end{equation}
where $\ell=e, \mu$ and the associated charges $Q_{\ell_L}, Q_{\ell_R}$ are defined in Eq.  (\ref{eq:charge2}).
Comparing with these models with LFU (lepton flavor universality), the FG2HDM has a potential to explain the  wrong sign
$\Delta a_\mu$ and $\Delta a_e$ due to the charge differences among various fermions.

\subsection{$R_{D^{(*)}}$ and $H^+$ }% and $B_c\to \tau \nu$}
\label{subsec:RD}

The observable  $R_{D^{(*)}}$ occurred in $B\to D^{(*)} \ell \nu$ decays are defined as
\begin{equation}
R_{D^{(*)}}=\frac{\mathcal{B}(B\to D^{(*)}\tau\bar{\nu})}{\mathcal{B}(B\to D^{(*)}\ell\bar{\nu})}\Big|_{\ell=e,\mu}.
\end{equation}
There are types of contributed new physics candidates, including exotic charged Higgs,  charge gauge bosons,
leptoquarks and so on.
In FG2HDM, the exotic charged Higgs $H^+$ is naturally accomondated. By integrating out the heavy scalar,
the quark level decay $b\to c\bar{\ell} \nu$ can be depicted by the following effective Hamiltonian
\begin{equation}
\mathcal{H}= C_{RL}^{\alpha} O_{RL}^{\alpha}+ C_{LL}^{\alpha} O_{LL}^{\alpha}
+C_{LR}^{\alpha} O_{LR}^{\alpha}+ C_{RR}^{\alpha} O_{RR}^{\alpha}
\end{equation}
with  four effective operators
\begin{align}
&O_{RL}^{\alpha k} = (\bar{c} \PR b)(\bar{\ell}_k \PL \nu_\alpha),\quad
O_{LL}^{\alpha k} = (\bar{c} \PL b)(\bar{\ell}_k \PL \nu_\alpha),\nonumber\\
&O_{LR}^{\alpha k} = (\bar{c} \PL b)(\bar{\ell}_k\PR \nu_\alpha),\quad
O_{RR}^{\alpha k} = (\bar{c} \PR b)(\bar{\ell}_k \PR \nu_\alpha),
\end{align}
and their corresponding Wilson coefficients 
\begin{align}
&C_{RL}^{\alpha k }=-\frac{2}{m_H^2 v^2} (V N_d)_{23} ( N_\ell)^\dagger_{\alpha k},\quad
C_{LL}^{\alpha k }=\frac{2}{m_H^2 v^2} (N_u V)_{23} ( N_\ell)^\dagger_{\alpha k},\\
&C_{LR}^{\alpha k }=-\frac{2}{m_H^2 v^2} (N_u V)_{23} (N_\nu )^\dagger_{\alpha k},\quad
C_{RR}^{\alpha k }=\frac{2}{m_H^2 v^2} (V N_d)_{23}  (N_\nu )^\dagger_{\alpha k}.\nonumber
\end{align}
where $V$ is CKM matrix, $m_H$ is the mass of charged Higgs $H^+$ and flavor index  $\alpha=e,\mu,\tau$ for neutrinos
and flavor index $k=1,2,3$ for charged leptons.
The two coefficients  $C_{LR, RR}^{\alpha k}$  are negligible since their sizes are proportional to neutrino mass,
leading to the two dominated contributions 
\begin{align}
& C_{RL}^{\tau 3}%=   \frac{2 m_b m_\tau V_{cb}}{m_H^2 v^2}
%\left[\left(1+\frac{1}{\tan^2\beta}\right) (V_{ub}^\dagger  +V_{cb}^\dagger
%+V_{tb}^\dagger)V_{tb} -1
%\right]
%\approx \frac{2m_b m_\tau}{m_H^2 v^2 \tan^2\beta} V_{cb}
\approx -2\sqrt{2}G_F V_{cb}\frac{m_b m_\tau}{m_H^2}\left(\frac{2}{ \tan^2\beta} +1 \right)
,\nonumber\\
&  C_{LL}^{\tau 3}=    -2\sqrt{2} G_F V_{cb} \frac{m_c m_\tau}{m_H^2}.
\label{eq:RDWC}
\end{align}
In particular, one can see in $C_{RL}^{\tau 3}$, 
the dependent behavior of $\tan\beta$ and $m_H$ changes dramatically comparing with 2HDM-II \cite{Hou:1992sy} and MSSM \cite{Buras:2002vd}.

Based on \cite{Fajfer:2012vx},  in FG2HDM, the  charged Higgs contribution to $R_D$ and $R_{D^*}$ can be further parameterized  as
\begin{align}
& R_D = R_D^{\rm SM} \left[ 1+1.5 {\rm Re} \left(\frac{C_{RL}^{\tau 3} +C_{LL}^{\tau 3}}
{C_{\rm SM}^{cb}}
\right) +1.0
\left| \frac{C_{RL}^{\tau 3} +C_{LL}^{\tau 3}}{C_{\rm SM}^{cb}} \right|^2\right],\\
& R_{D^{*}} = R_{D^{*}} ^{\rm SM} \left[ 1+0.12 {\rm Re} \left(\frac{C_{RL}^{\tau 3} -C_{LL}^{\tau 3}}
{C_{\rm SM}^{cb}}
\right) +0.05
\left| \frac{C_{RL}^{\tau 3} -C_{LL}^{\tau 3}}{C_{\rm SM}^{cb}} \right|^2
\right],\nonumber
\end{align}
where 
%\begin{equation}
$C_{\rm SM}^{cb} = 2\sqrt{2}G_F V_{cb}$. 
%\end{equation}

%\newpage

%\section{$B_c\to \tau^-\bar{\nu}$}
A correlated process to $B\to D^{(*)}\bar{\ell}\nu$ is $B_c\to \tau \nu$.
Incorporating scalar operator contribution, one obtains branching ratio of $B_c\to \tau^-\bar{\nu}$ 
\cite{Huang:2018nnq, Akeroyd:2017mhr}
in FG2HDM,
\begin{equation}
\mathcal{B} (B_c\to \tau^-\bar{\nu})= \frac{1}{8\pi}\tau_{B_c}
G_F^2|V_{cb}|^2 m_{B_c}m_\tau^2 f_{B_c}^2\left(1-\frac{m_\tau^2}{m_{B_c}^2}\right)^2
\left| 1+ \frac{m_{B_c}^2}{(m_b+m_c)m_\tau}C_P\right|^2
\end{equation}
where $C_P=\left(C_{RL}^{\tau 3}-C_{LL}^{\tau 3}\right)/C_{\rm{SM}}^{cb}$.
A numerical analysis to solution space will be carried on in the following Sec. \ref{sec:num}.

%\newpage
\section{Numerical Analysis}
\label{sec:num}
\subsection{Experimental status and inputs}
\label{subsec:exp}

There have been continuous updates for the measurements of $R_K$  
%\cite{Aaij:2014ora, Aaij:2019wad, Abdesselam:2019lab} 
and $R_{K^*}$ by LHCb, Belle, CMS and ATLAS. The latest measurement, given by LHCb during Moriond 2021,
shows the $3.1\sigma$  deviation  in $R_K$ (see Eq. (\ref{eq:RK})) and 
confirms the tension between SM. To explore the dynamics in high energy, we make use of global fitting results, which rely on both the experimental data as well as the choices of fitting basis. The latest fitting results, including the  LHCb
new $R_K$ measurement, are presented in Moriond QCD 2021\cite{Alguero:Moriond2021}.

Based on 2019 data, the global fit works done by several independent groups \cite{Arbey:2019duh,Aebischer:2019mlg,Alguero:2019ptt} are consistent well in the following facts: i) large and negative 
$\delta C_9^\mu (\textrm{best\, fit} \sim -1)$
%\footnote{
%Note here the convention for Wilson coefficients $C_{9,10}$ corresponds to
%operator definitions $\mathcal{O}_9= \frac{\alpha_{\textrm{em}}}{4\pi} (\bar{s}\gamma_\mu \PL b)
%(\bar{\ell}\gamma^\mu \ell)$,
%$\mathcal{O}_{10}= \frac{\alpha_{\textrm{em}}}{4\pi} (\bar{s}\gamma_\mu \PL b)
%(\bar{\ell}\gamma^\mu\gamma_5 \ell)$, hence $\Delta C_{9,10}^\ell = \frac{\alpha_{\textrm{em}}}{4\pi} \delta C_{9,10}^\ell$
%}
 , ii) relative small and positive $\Delta C_{10}^\mu ( \textrm{best\, fit} \sim 0.5)$. Two other parameters $\Delta C_{9,10}^e$ were only contained in the analysis of \cite{Aebischer:2019mlg,Alguero:2019ptt}, sharing the common features: i) positive and relative large $\Delta C_9^e (\textrm{best\, fit} \sim 0.8)$
 and ii) negative and relative large $\Delta C_{10}^e (\textrm{best\, fit} \sim -0.78)$. In the new fit of \cite{Alguero:Moriond2021} (2D fit), $\Delta C_9^e$ is included compared with
the previous work \cite{Alguero:2019ptt} and changes dramatically from \cite{Arbey:2019duh,Aebischer:2019mlg}:
the sign of central value has been flipped. On the other hand, $\Delta C_{10}$ is still untouched. Hence a more complete global fit is highly anticipated.  
In current work, we  mainly adopt the central values of 2D and 6D fits in \cite{Alguero:Moriond2021}, and  conjecture  the
untouched $\Delta C_{10}^e$ combining the results in \cite{Arbey:2019duh,Aebischer:2019mlg} based on old data, 
giving
 \begin{align}
& \Delta C_9^\mu= -1.21\pm 0.20   ,\qquad \Delta C_9^e=   -0.40\pm0.40, \\
& \Delta C_{10}^\mu= 0.15 \pm 0.20   ,\qquad \Delta C_{10}^e= -0.78\pm0.40.      \nonumber
\end{align}
In particular, we have allowed more tolerant errors.

For the  experimental values $R_{D}$ and $R_{D^{*}}$, we adopt  world averages from the 
heavy flavor averaging group (HFLAV) \cite{Amhis:2019ckw}
\begin{equation}
R_D=0.340\pm 0.027\pm 0.013,\qquad
R_{D^*}=0.295\pm 0.011\pm 0.008,
\end{equation}
which are based on measurements from BaBar, Belle and LHCb.
The corresponding SM predictions are known with high precision, reads 
\begin{equation}
R_D^{\rm{SM}}=0.299\pm 0.003,\qquad
R_{D^*}^{\rm{SM}}=0.258\pm 0.005,
\end{equation}
which is also quoted from HFLAV \cite{Amhis:2019ckw}.

The lifetime of $B_c$ meson, we adopt the latest PDG value \cite{Zyla:2020zbs}
\begin{equation}
\tau_{B_c} = 0.510\pm 0.009  \; {\rm ps} 
\end{equation}
The decay mode $B_c\to \tau^-\nu$ has not been measured. Here we take 3 conjectures
(see also \cite{Akeroyd:2017mhr}), 
\begin{equation}
\mathcal{B} (B_c\to \tau^-\bar{\nu}) \leq 30\%;\; 20\%;\; 10\%.
\end{equation}
for convenience. 

 %\subsubsection*{$(g-2)_\mu$}
 Improvements of muon AMDM are made  due to efforts from both the theoretical and experimental sides.
 The latest calculation in SM, including $\mathcal{O}(\alpha^5)$ QED correction, electroweak 
 correction, NNLO hadronic vacuum polarization (HPV) as well as Hadronic Light-by-Light (HLbL) contributions, is summarized in the review \cite{Aoyama:2020ynm} , giving
 \begin{equation}
 a_\mu^{\rm{SM}}=(116\, 591\, 810\pm 43 ) \times 10^{-11}.
 \end{equation}
It differs  the Brookhaven
 measurement \cite{Bennett:2006fi}
 %\begin{equation}
 $
 a_\mu^{\rm{BNL}} = (116\, 592\, 089\pm 63 )\times 10^{-11}
 $
 %\end{equation}
 by
 %\begin{equation}
 $
 \Delta a_\mu := a_\mu^{\rm{BNL}}-a_\mu^{\rm{SM}}=(279\pm 76)\times 10^{-11},
 $
 %\end{equation}
corresponding to a $3.7\sigma$ discrepancy. Recently, the Muon g-2 Experiment at Fermilab released
their first result %\cite{Polly:muongm2} 
\cite{PhysRevLett.126.141801}
after nearly 20 years from Brookhaven's result,
%\begin{equation}
 $
 a_\mu^{\rm{FNAL}} = (116\, 592\, 040\pm 54 )\times 10^{-11}
$
%\end{equation}
leading to the latest 
%\begin{equation}
 $
 \Delta a_\mu^{\rm{FNAL}} = ( 230\pm 69 )\times 10^{-11},
$
%\end{equation}
corresponding to a $3.3 \sigma$ discrepancy, which 
confirms the existence of a tension and strengthens the evidence of new physics.
Then the experimental average, by combining BNL and FNAL results together, is given 
\cite{PhysRevLett.126.141801} as
\begin{equation}
a_\mu^{\rm{exp}}=(116\, 592\, 061\pm 41 )\times 10^{-11}
\end{equation}
and hence the deviation is 
\begin{equation}
\Delta a_\mu^{\rm{2021}} = ( 251\pm 59 )\times 10^{-11},
\label{eq:gm22021}
\end{equation}
with a $4.2 \sigma$ significance. In the following numerical calculation, we will take the new combined 
result Eq.(\ref{eq:gm22021}) as the input.

 %\subsubsection*{$(g-2)_e$}
Recently an improved measurement \cite{Parker:2018vye} of the fine-structure
constant $\alpha$ toward a deviation in the electron AMDM from theoretical prediction 
\begin{equation}
\Delta a_e = -(8.7\pm 3.6) \times 10^{-13},
\end{equation}
corresponding to a negative $2.4\sigma$ discrepancy.
It is worthy pointing out that the sign of $\Delta a_e$ differs the one of $\Delta a_\mu$.
%\subsubsection*{$\mu\to e \gamma$}
%The  MEG collaboration gave an upper bound for the LFV process $\mu\to e \gamma$ in 2016
%\cite{TheMEG:2016wtm}
%\begin{equation}
%\mathcal{B}(\mu^+\to e^+ \gamma)<4.2\times 10^{-13}
%\end{equation}
%at $90\%$ confidence level.

Other input parameters are summarized in the 
 in Table \ref{tab:input}.
\begin{table}
	\caption{Input parameters used in the numerical analysis.}
	\label{tab:input}
	\begin{center}
			\begin{tabular}[b]{cc}
				\hline\hline
				Parameters & Values  \\
				\hline
				$V_{cb}$ & %$41.0\pm1.4\times10^{-3}$
				 $(42.2\pm0.8)\times10^{-3}$ \\
				$V_{ts}$& %$38.8\pm1.1\times10^{-3}$ 
				$(39.4\pm2.3)\times10^{-3}$ \\
				$V_{tb}$& %$1.013\pm0.030$ 
				$1.019\pm0.025$ \\
				$G_{F}$&$1.1663787\times10^{-5}\textrm{~GeV}^{-2}$\\
				$m_{b}$&$4.18^{+0.03}_{-0.02}\textrm{~GeV}$\\
				$m_{\tau}$&$1776.86\pm0.12\textrm{~MeV}$\\
				$m_{c}$&$1.27\pm0.02\textrm{~GeV}$\\
				$m_{W}$&$80.379\pm0.012\textrm{~GeV}$\\
				$m_{Z^0}$&$91.1876\pm0.0021\textrm{~GeV}$\\
				$m_{e}$&$0.5109989461\pm3.1\times10^{-9}\textrm{~Mev}$\\
				$m_{\mu}$&$105.6583745\pm2.4\times10^{-6}\textrm{~Mev}$\\
				$\alpha(m_{W})$&$\frac{1}{128}$\\
				$m_{B_c}$&$6274.9\pm0.8\textrm{~MeV}$\\
				$f_{B_c}$&$0.434\textrm{~GeV}$\\
				$\tau_{B_c}$&$0.510\pm0.009~\textrm{ps}$\\
				\hline
			\end{tabular}
		\end{center}
\end{table}

\subsection{Numerical results}

\begin{figure}[h]
\begin{center}
$$
\begin{array}{cc}
\includegraphics[width=0.45\textwidth]{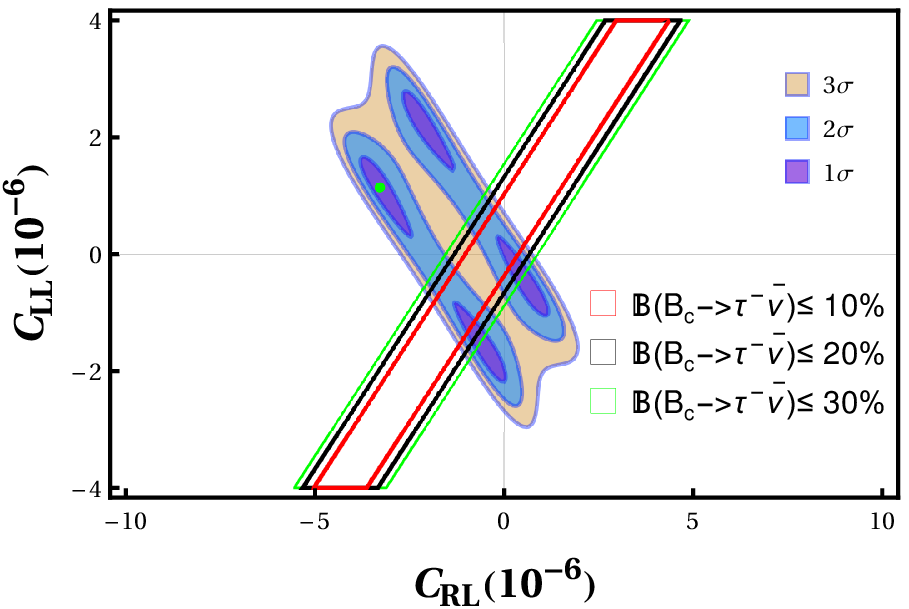}&
\includegraphics[width=0.5\textwidth]{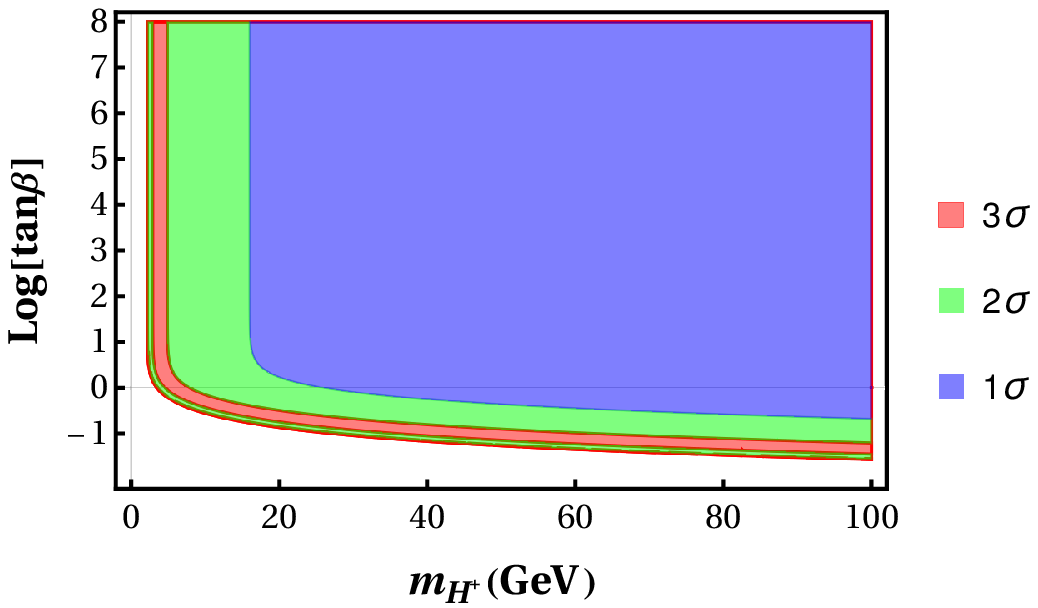}
%&\includegraphics[width=0.3\textwidth]{fig/fig2c.eps}
\\
(a) & (b) 
%& (c)
\end{array}
$$
\caption{The  allowed parameter space from $b\to c \ell \nu$: (a)
allowed regions for Wilson Coefficients of scalar operators by $R_D^{(*)}$ 
%within $1\sigma $, $2\sigma $ and $3\sigma $.
associated with
conjectures of $B_c\to \tau \nu$;
(b) the allowed parameter space purely from $R_{D^{(*)}}$ in FG2HDM
%; (c) the comparison of boundaries of $3\sigma$ $R_{D^{(*)}}$ and $30\%$ upper bound of $B_c\to \tau\nu$
.}
\label{fig:RDandBc}
\end{center}
\end{figure}

The priority here in FG2HDM is to find a solution space after introducing three additional particles in an economic way. 
We take two U(1) charges\footnote{
Generally speaking, there are two free U(1) charges in FG2HDM. In the scenario shown in Eq.(\ref{eq:rela-phi}),
imposing the decoupling limit of $Z'$, one degree of freedom can
be eliminated.
}
, $\tan\beta$, $m_{H^+}$ and $g'/m_{Z'}$ as free parameters in the following numerical calculation. 
Since the relying parameters are uncorrelated so far\footnote{ A more comprehensive analysis
of FG2HDM including more observables is in progress and to be shown shortly.}
, we hence carry out the calculation of 
$R_{D^{(*)}}$ and $R_{K^{(*)}}$ separately. 
It is understandable that both anomalies can be accommodated in FG2HDM once their corresponding 
solution space is found.

The parameter space of $R_{D^{(*)}}$ associated with $B_c\to \tau\nu$ are presented  Fig.\ref{fig:RDandBc}.
In a more general model with scalar operators, $R_{D^{(*)}}$ has already put  strong constraints, by shown the $1\sim 3\sigma$
allowed regions in Fig.\ref{fig:RDandBc}(a), one can see 4 allowed areas at $1\sigma$ level  in $C_{RL}-C_{LL}$ space 
($C_{RL,LL}$ are the coefficients in Eq.(\ref{eq:RDWC})). However, even with a very loose upper bound (say $30\%$), $B_c\to\tau\nu$
helps to exclude half of the regions. In the case of FG2HDM, the situation is quite friendly. As shown in Fig.\ref{fig:RDandBc}(b), most 
area in $\log(\tan\beta) - m_{H^+}$ space are allowed
 by $R_{D^{(*)}}$, especially for the large $\tan\beta$ region. The restriction from $B_c\to \tau\nu$ in $\log(\tan\beta) - m_{H^+}$ space
 is weak thus we do not show its effect in Fig.\ref{fig:RDandBc}(b).
 In fact,  we have compared the the boundaries of $3\sigma$ $R_{D^{(*)}}$ and $30\%$ upper bound of  $B_c\to \tau\nu$ and find they are close to each other.  
 %in Fig.\ref{fig:RDandBc}(c),
 %we can see it is reasonable for such a neglect in Fig.\ref{fig:RDandBc}(b).

 \begin{figure}[t]
\begin{center}
$$
\begin{array}{cc}
\includegraphics[width=0.45\textwidth]{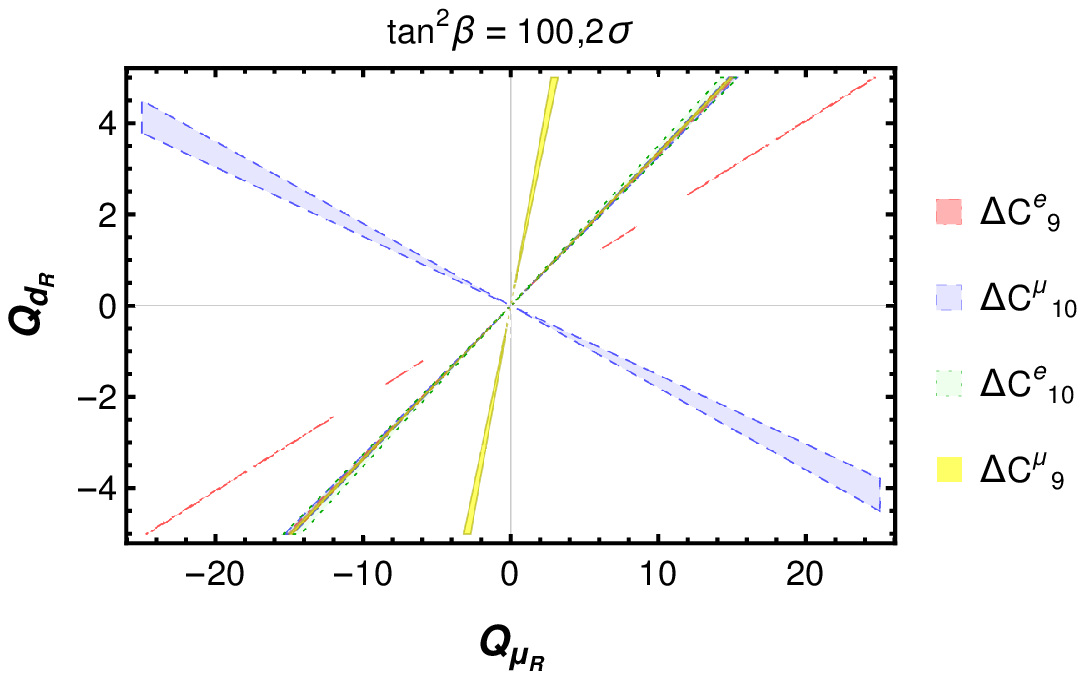}&
\includegraphics[width=0.45\textwidth]{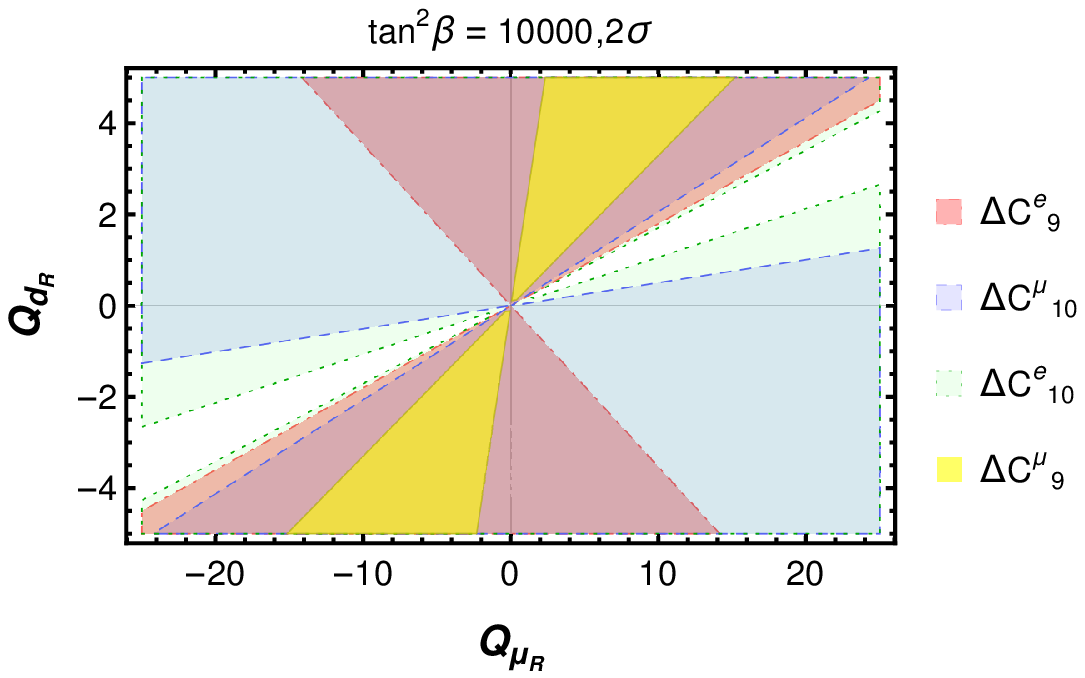}\\
(a) & (b) \\
\includegraphics[width=0.45\textwidth]{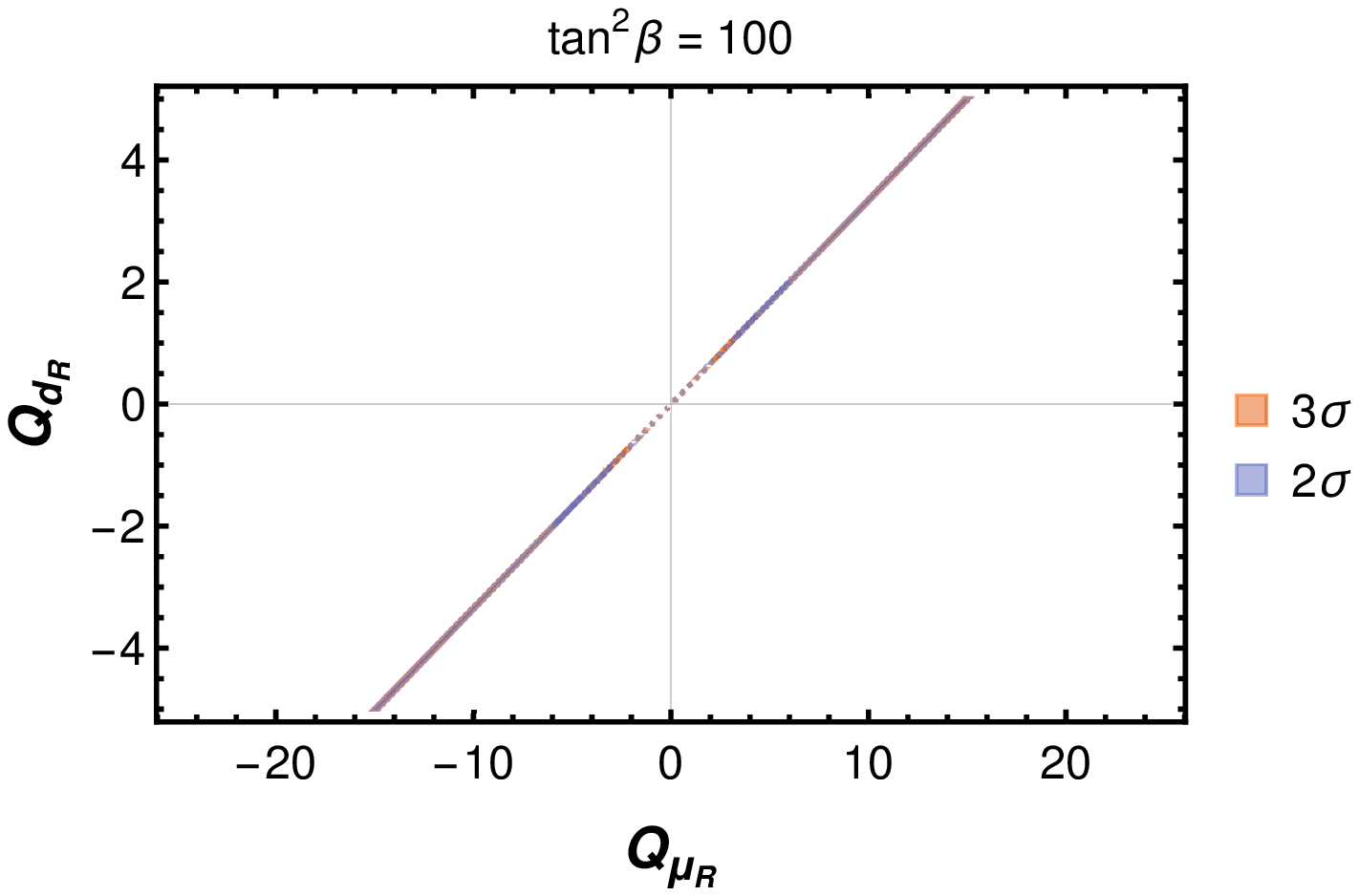}&
\includegraphics[width=0.45\textwidth]{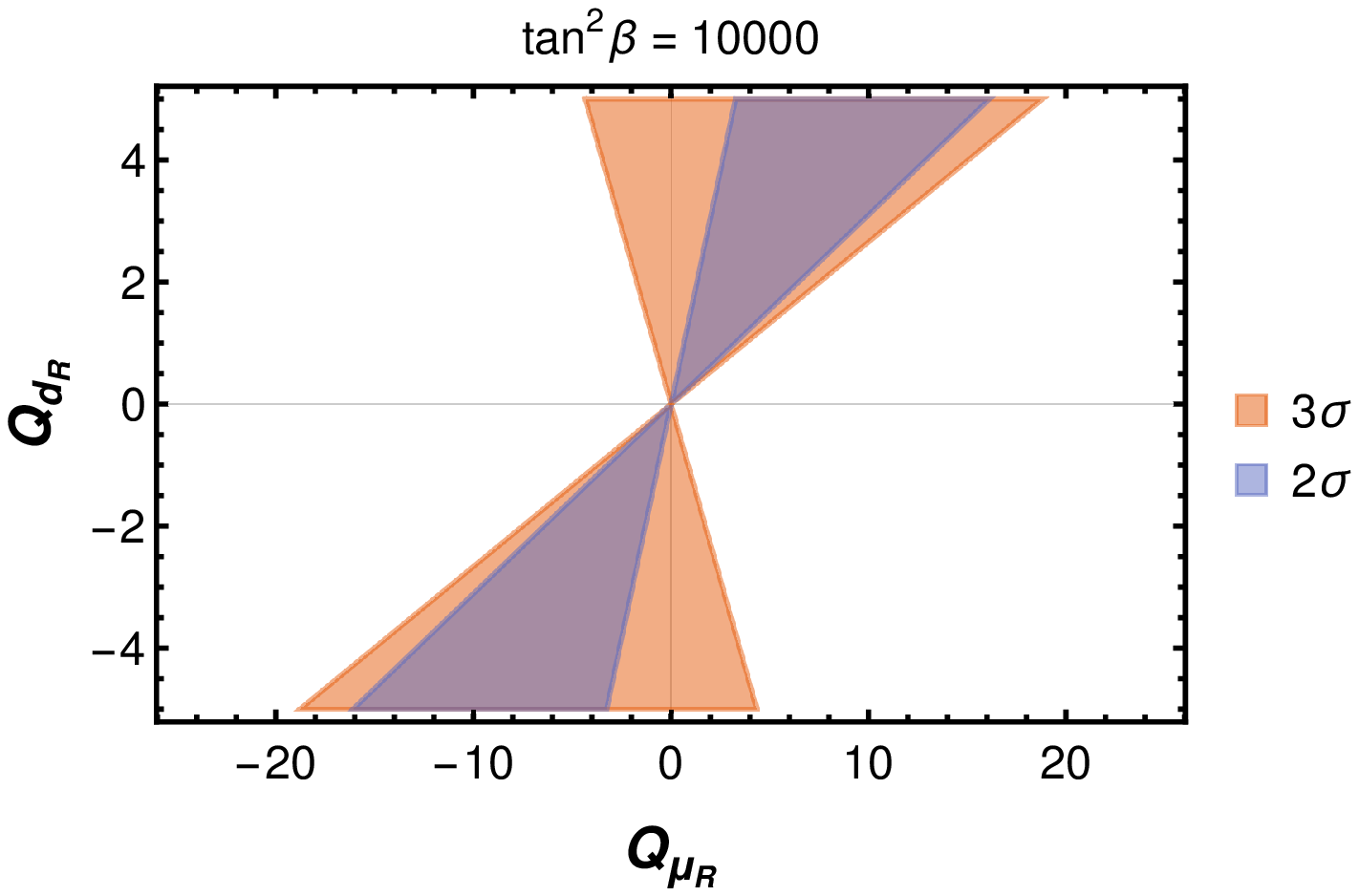}\\
(c) & (d) \\
\end{array}
$$
\caption{The allowed free U(1) charges  by $R_{K^{(*)}}$ at $2\sigma$ with different choices of
$\tan\beta$:
(a)  various $\Delta C_i^\ell$ allowed regions with $\tan^2\beta=10^2$; 
(b) various $\Delta C_i^\ell$ allowed regions  with $\tan^2\beta=10^4$;
(c) the allowed region combined all $\Delta C_i^\ell$ with $\tan^2\beta=10^2$
and 
(d) the allowed region combined all $\Delta C_i^\ell$ with $\tan^2\beta=10^4$
.}
\label{fig:RK}
\end{center}
\end{figure}

The physical regions of U(1) charges, constrained by $R_{K^{(*)}}$ are illustrated in Fig.\ref{fig:RK}.  As shown in Fig.\ref{fig:RDandBc}(b),
large $\tan\beta$ is favored in $b\to c\ell\nu$. To show the $\tan\beta$ dependence in $b\to s\ell\bar{\ell}$, two scenarios with $\tan^2\beta=100$
and $\tan^2\beta=10000$ are chosen. Hence one may get the impression that the larger $\tan\beta$ is taken, the wider solution space can
be found. We show in upper row the detailed regions allowed by four Wilson coefficients. Taking Fig.\ref{fig:RK}(b) as an example, 
the constraints given by 
$\Delta C_{10}^e$ is loose. It is reasonable as we have made  a relatively loose  conjecture on it based on old data. On the contrary, the region allowed by $\Delta C_9^\mu$ is narrowest as there are more fit results and hence more precise constraints putting on it. Then we show the survived space combined all the constraints to the 4 Wilson coefficients in Fig.\ref{fig:RK}(c) and (d) in $2\sigma$ and $3\sigma$ level.

\begin{figure}[t]
\begin{center}
$$
\begin{array}{cc}
\includegraphics[width=0.45\textwidth]{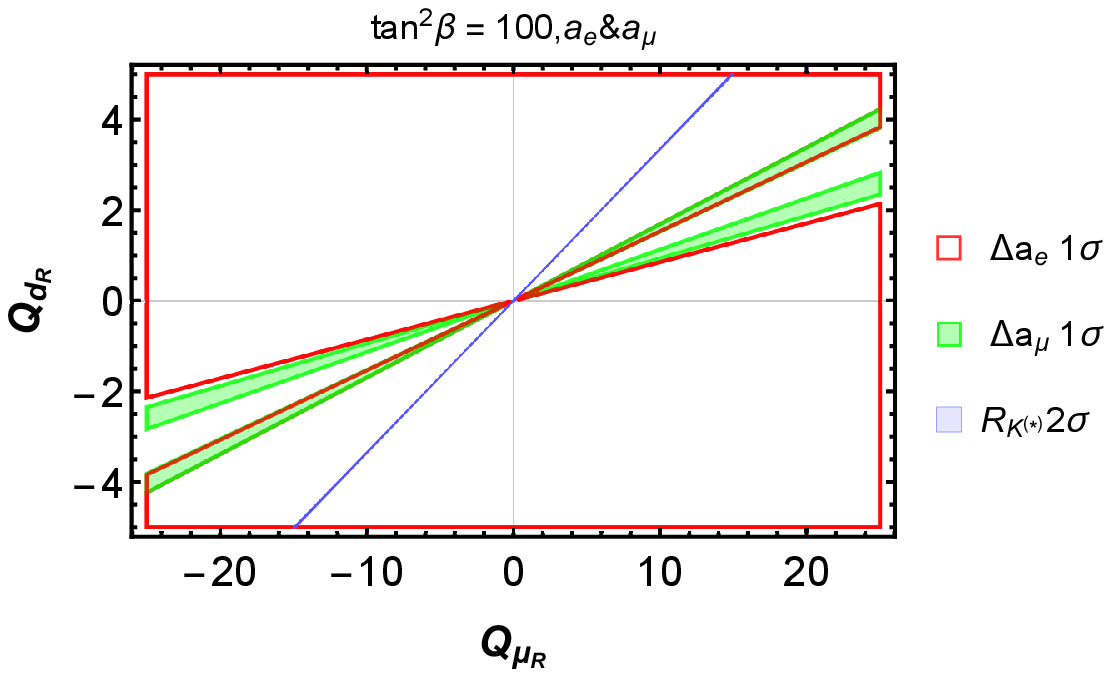}&
\includegraphics[width=0.45\textwidth]{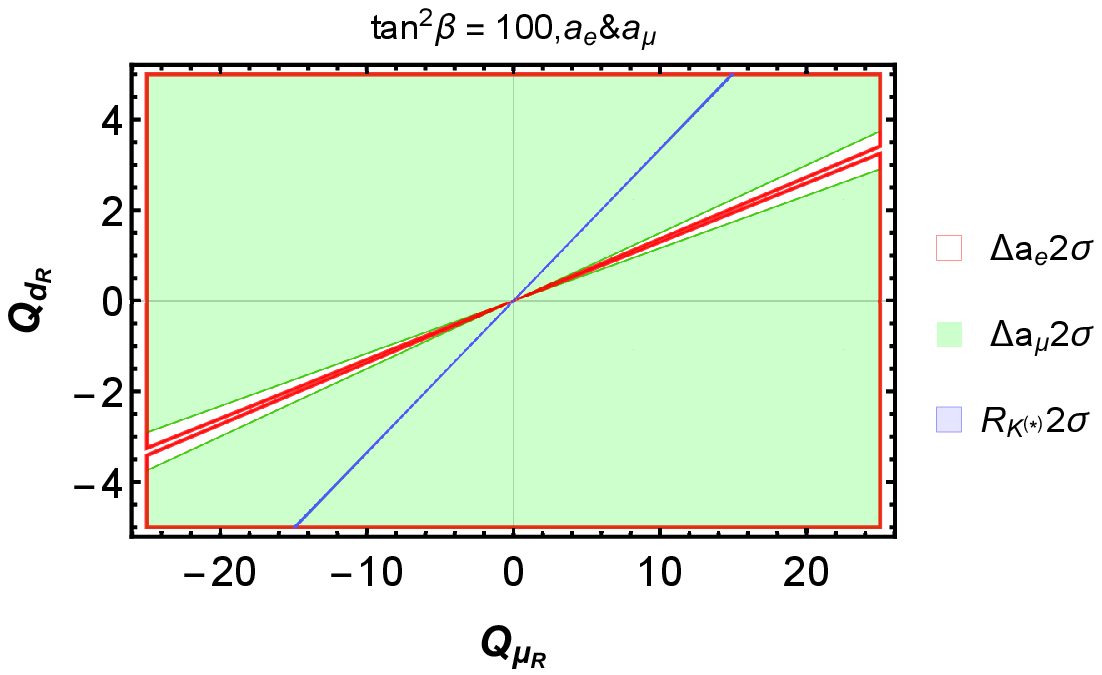}\\
(a) & (b) 
\end{array}
$$
\caption{The allowed free U(1) charges  further discriminated by $\Delta a_\ell$:
(a)  $2\sigma$ allowed region by $\Delta a_e$ and $\Delta a_\mu$ with $\tan^2\beta=10^2$; 
(b) $2\sigma$ allowed region by $\Delta a_e$ and $\Delta a_\mu$  with $\tan^2\beta=10^4$.}
\label{fig:gm2}
\end{center}
\end{figure}

We show the behaviors of  AMDM parameter space, especially making use of the new muon AMDM measurement, in Fig.\ref{fig:gm2}.
Though the dependence of $\tan\beta$ is not linear, typically one may observe small $\tan\beta$ is somehow favored by AMDM. 
To extract the main features, we fix $\tan^2\beta=100$ in Fig.\ref{fig:gm2} as an illustration. From the $1\sigma$ allowed region by $\Delta a_\mu$
and $\Delta a_e$ shown in Fig.\ref{fig:gm2}(a), one may find: i) $\Delta a_\mu$ in general has a more narrow range than $\Delta a_e$; 
ii) the overlap range of $\Delta a_e$ and $\Delta a_\mu$ is largely reduced due to their opposite trend shown in  Fig.\ref{fig:gm2}(a), originated
from their sign difference; iii) the survived parameter space combing both AMDMs almost has no overlap with the allowed area from $R_{K^{(*)}}$,
except the trivial solution around the origin. However, from Fig.\ref{fig:gm2}(b) the situation changes if $2\sigma$ error of AMDMs is allowed. In this
case, $\Delta a_e$ gives almost no constraints while the constraint from $\Delta a_\mu$ is stronger, but still fills most of the presented area and entirely contains the $2\sigma$ allowed region from $R_{K^{(*)}}$.

Some points are summarize  as follows:
\begin{itemize}
\item There is plenty of solution space for $R_{D^{(*)}}$.
\item It is challenging to obtain $1\sigma$ allowed region from $R_{K^{(*)}}$, but there is rich
$2\sigma$ solution, which is large $\tan\beta$ favorite.
\item Though the solution exists purely from $1\sigma$ $\Delta a_\mu$ and $\Delta a_e$, but $2\sigma$ $R_{K^{(*)}}$
solution almost kills $1\sigma$ $\Delta a_\ell$ solution.
\item The $2\sigma$ solution exists combing the latest $R_{K^{(*)}}$ and $\Delta a_\ell$ in FG2HDM.

\end{itemize} 

%we are eager to see whether the solution space related to $Z'$ can be
%further restricted based on $R_{K^{(*)}}$. As presented in Fig.\ref{fig:gm2}(a), $\Delta a_e$ has a much wider solution region, containg
%all the solution of $2\sigma$ $R_{K^{(*)}}$, and has a minor overlap with $\Delta a_\mu$, indicating the common solution located around
%the origin. The trend keeps, though with larger $\tan\beta$ shown in Fig.\ref{fig:gm2}(b). 
%With the discrimination of muon AMDM, the allowed parameter space of $R_{K^{(*)}}$ has been reduced to a narrow region.
%However, if we loose the constraint from muon AMDM to $2\sigma$, the common region  entirely governs by $R_{K^{(*)}}$.

%\begin{figure}[t]
%\begin{center}
%$$
%\begin{array}{cc}
%\includegraphics[width=0.45\textwidth]{fig/fig3a.eps}&
%\includegraphics[width=0.45\textwidth]{fig/fig3c.eps}\\
%(a) & (b) \\
%\end{array}
%$$
%\includegraphics[width=8cm]{leptonic.eps}
%\caption{Allowed parameter space in FG2HDM by $R_D^{(*)}$ anomalies
%within $1\sigma $, $2\sigma $ and $3\sigma $.
%and
%conjectures of $B_c\to \tau \nu$.}
%\label{fig:RK2}
%\end{center}
%\end{figure}

\section{Conclusion and Outlook}
\label{sec:con}

The Yukawa interaction and Higgs sector are naturally connected via spontaneously symmetry breaking. To get rid of
the redundancy in Yukawa coupling of 2HDM-III, a specific U(1) flavor symmetry is introduced leading to FG2HDM.
This symmetry brings different charges to the two Higgs doublets and hence forbids the $\lambda_5$ and $m_{12}$ terms
in scalar potential. No physical pseudoscalar turns up and there are only three additional particles ($H^+, H^0$ and $Z'$) adding 
to SM particle spectrum. 

The exotic neutral scalar and exotic gauge boson both can mediate flavor-changing current when they interact with down-type quark.
In this work, we particularly investigate the role of $H^+$ and $Z'$ in the interpretation of the recent flavor anomalies.
We make use of $Z'$ to generate new contribution to $b\to s\ell\bar{\ell}$, which helps to explain $R_{K^{(*)}}$ anomaly at $2\sigma$ level. 
There is a tension between $2\sigma$ $R_{K^{(*)}}$ and $1\sigma$ $\Delta a_\ell$, combing 
the new released FNAL muon AMDM and recently improved electron AMDM.
But the solution is safe if both $R_{K^{(*)}}$ and $\Delta a_\ell$ allow $2\sigma$ error.
There are plenty of rooms on $(m_{H^+},\tan\beta)$ plane to provide solutions to
$R_{D^{(*)}}$ anomalies. The rich parameter space, especially the one for explaining $R_{D^{(*)}}$ anomalies, will
be confronted with more examinations in next step.

Nevertheless, currently we may conclude that  in FG2HDM
a solution indeed exists at $2\sigma$ level for the tensions  in $R_{D^{(*)}}, R_{K^{(*)}}$ and $\Delta a_{\mu}, \Delta a_e$.

%The ongoing examination and investigation of FG2HDM, including to 

%In next step, we will go further to examine the remaining parameter space. Apparently, taking Fig.\ref{fig:RDandBc}(b) as example, 
%current constraint is not suf
%the constraint on 

\begin{acknowledgments}

This research  is supported by NSFC under Grant No. U1932104.
\end{acknowledgments}

\newpage

\appendix

\section{Some details of the model}
\label{app:details}

\subsection{Quantum numbers for flavor gauge symmetry}
\label{app:symmetry}
The nice texture of Yukawa matrices in Eq. (\ref{eq:quarkYukawa}) and Eq. (\ref{eq:lepYukawa}) 
can be 
guaranteed by  the subtile symmetry introduced by an extra U(1) transformation.
Under this extended U(1) group,  the  behaviors of all the relavent fields are given as follows,
\begin{equation}
\phi \to \phi'= e^{i\theta X_\phi} \phi,
\end{equation}
where the U(1) charges to keep the above Yukawa structures are chosen to be
\begin{align}
& X_{Q_L}=\frac12 \left(
\begin{array}{ccc}
Q_{u_R}+Q_{d_R} & & \\
 & Q_{u_R}+ Q_{d_R} &  \\
 & & Q_{t_R}+Q_{d_R}\end{array}
\right), \label{eq:charge}\\%\nonumber\\
& X_{u_R}= \left(
\begin{array}{ccc}
Q_{u_R} & & \\
 &Q_{u_R} &  \\
 & &Q_{t_R}\end{array}
\right),\qquad
 X_{d_R}= \left(
\begin{array}{ccc}
Q_{d_R} & & \\
 & Q_{d_R} &  \\
 & & Q_{d_R}\end{array}
\right),\nonumber\\
& X_\Phi=\frac12\left(\begin{array}{cc}
 Q_{u_R}-Q_{d_R} &\\
  & Q_{t_R}-Q_{d_R}
  \end{array}\right),
  \nonumber\\
  & X_{L_L}=\left(
\begin{array}{ccc}
Q_{e_L} & & \\
 & Q_{\mu_L} &  \\
 & & Q_{\tau_L}\end{array}
\right),\nonumber\\
& X_{\ell_R}=\left(
\begin{array}{ccc}
Q_{e_R} & & \\
 & Q_{\mu_R} &  \\
 & & Q_{\tau_R}\end{array}
\right),\qquad X_{\nu_R}=0. \nonumber
\end{align}
However, the charges are not that free as they should satisfy anomaly cancellation conditions.
Imposing the anomaly cancellation, we established relations among these charges as
%Anomaly cancellation contions give the following relations
\begin{align}
& Q_{u_R}=-Q_{d_R}-\frac13 Q_{\mu_R},\quad
 Q_{t_R}=-4Q_{d_R}+\frac23 Q_{\mu_R},\nonumber\\
& Q_{\tau_L}=Q_{d_R}+\frac16 Q_{\mu_R},\quad\quad
Q_{\mu_L}=-Q_{d_R}+\frac56 Q_{\mu_R},\quad
 Q_{e_L}=\frac92 Q_{d_R}-Q_{\mu_R},\nonumber\\
& Q_{\tau_R}=2Q_{d_R}+\frac13 Q_{\mu_R},\quad\;
 Q_{e_R}=7Q_{d_R}-\frac43 Q_{\mu_R}\label{eq:charge2}
\end{align}
leaving only two degrees of freedom, denoted as
$Q_{d_R}, Q_{\mu_R}$. Our result here is consistent with the one in
\cite{Celis:2015ara} by permuting  $e$ and $\tau$.

\subsection{Interactions among scalars}
\label{app:scalar}
In Sec.\ref{subsec:scalar}, the detailed scalar interactions have been given.
Here, we further provide the exact 
coefficients among them, giving
\begin{equation}
	\begin{aligned}
		\lambda_{h^3}&=\frac32 v [2\lambda_{1}\cos\beta\cos^3\alpha+2\lambda_{2}\sin\beta\sin^3\alpha+\lambda_{34}\sin2\alpha\sin(\alpha+\beta)]\\
		\lambda_{h^2H^{0}}&=\frac14 v \{12\lambda_{2}\sin\beta\cos\alpha\sin^2\alpha-12\lambda_{1}\cos\beta\sin\alpha\cos^2\alpha+\lambda_{34}[\sin(\beta-\alpha)+3\sin(3\alpha+\beta)]\}\\
		\lambda_{hH^{0^2}}&=\frac14 v \{12\lambda_{1}\cos\beta\cos\alpha\sin^2\alpha+12\lambda_{2}\sin\beta\sin\alpha\cos^2\alpha+\lambda_{34}[\cos(\beta-\alpha)+3\cos(3\alpha+\beta)]\}\\
		\lambda_{H^{0^3}}&=\frac32 v [2\lambda_{2}\sin\beta\cos^3\alpha-2\lambda_{1}\cos\beta\sin^3\alpha-\lambda_{34}\sin2\alpha\cos(\alpha+\beta)]\\
		\lambda_{hH^+H^-}&=\frac14 v \{2\sin2\beta[\lambda_{2}\cos\beta\sin\alpha+\lambda_{1}\sin\beta\cos\alpha]+\lambda_{34}\cos(\alpha+3\beta)+(3\lambda_3-\lambda_4)\cos(\beta-\alpha)\}\\
		\lambda_{H^0H^+H^-}&=\frac14 v \{2\sin2\beta[\lambda_{2}\cos\beta\cos\alpha-\lambda_{1}\sin\beta\sin\alpha]-\lambda_{34}\sin(\alpha+3\beta)+(3\lambda_3-\lambda_4)\sin(\beta-\alpha)\}\\
		\lambda_{h^4}&=\frac34\sin^2 2\alpha[\lambda_{1}\cot^2 \alpha+\lambda_{2}\tan^2 \alpha+2\lambda_{34}]\\
		\lambda_{h^3H^0}&=\frac32 \sin 2\alpha[(\lambda_{34}-\lambda_{1})\cos^2\alpha+(\lambda_{2}-\lambda_{34})\sin^2\alpha]\\
		\lambda_{h^2H^{0^2}}&=\frac{1}{4}[(3\lambda_{1}+3\lambda_{2}-2\lambda_{34})\sin^2 2\alpha+4\lambda_{34}\cos^2 2\alpha]\\
		\lambda_{hH^{0^3}}&=\frac32 \sin 2\alpha[(\lambda_{2}-\lambda_{34})\cos^2\alpha+(\lambda_{34}-\lambda_{1})\sin^2\alpha]\\
		\lambda_{H^{0^4}}&=\frac34\sin^2 2\alpha[\lambda_{1}\tan^2\alpha+\lambda_{2}\cot^2 \alpha+2\lambda_{34}]\\
		\lambda_{h^2H^+H^-}&=\frac14 \sin2\alpha\sin2\beta[\lambda_{1}\tan\beta\cot\alpha+\lambda_{2}\tan\alpha\cot\beta+\lambda_3(\cot\alpha\cot\beta+\tan\alpha\tan\beta)-2\lambda_4]\\
		\lambda_{hH^0H^+H^-}&=\frac14 \sin2\beta\sin2\alpha[(\lambda_3-\lambda_{1})\tan\beta+(\lambda_{2}-\lambda_3)\cot\beta-2\lambda_4\cot2\alpha]\\
		\lambda_{H^{0^2}H^+H^-}&=\frac14 \sin2\alpha\sin2\beta[\lambda_{1}\tan\alpha\tan\beta+\lambda_{2}\cot\alpha\cot\beta+\lambda_3(\tan\beta\cot\alpha+\tan\alpha\cot\beta)+2\lambda_4]\\
		\lambda_{H^+H^-H^+H^-}&=\frac12 \sin^2 2\beta[\lambda_{1}\tan^2\beta+\lambda_{2}\cot^2\beta+2\lambda_{34}],
	\end{aligned}
\end{equation}
which will be helpful in the Higgs phenomenology studies.

%\appendix
%\section{BGL model}

%\newpage

%\appendix

%\section{The charge assignment of new gauge group $U(1)'$}

%\newpage

%\acknowledgments
%F. Xu acknowledges the support by NSFC  under Grant No. 11605076.

%\appendix

%\section{Appendix 1}
%\label{sec:app}

\bibliographystyle{app/JHEP}
\bibliography{app/reference}

%\begin{thebibliography}{99}
 %asd
%\end{thebibliography}
 
\end{document}